\documentclass[%
 reprint,
nofootinbib,
 amsmath,amssymb,
 aps,
]{revtex4-2}

\usepackage{graphicx}
\usepackage{dcolumn}
\usepackage{bm}

\usepackage{subcaption}

\newcommand{\jcap}{Journal of Cosmology and Astroparticle Physics}
\newcommand{\aap}{Astron. Astrophys.}
\newcommand{\mnras}{Mon. Not. R. Astron. Soc.}
\newcommand{\apjl}{Astrophys. J. Lett.}
\newcommand{\physrep}{Physics Reports}

\begin{document}

\preprint{APS/123-QED}

\title{Nonminimally coupled ultralight axions as cold dark matter}

\author{Kishan Sankharva}
\email{khsankha@syr.edu}
\affiliation{Department of Physics, Syracuse University, Syracuse, NY 13244, U.S.A.}
\author{Shiv Sethi}%
 \email{sethi@rri.res.in}
\affiliation{Raman Research Institute, Bengaluru, Karnataka 560080, India}

\date{\today}

\begin{abstract}

We consider a nonminimally coupled scalar field as a potential cold dark matter candidate.  These
models are natural extensions of the ultralight axion models which are  based on minimally coupled
scalar fields. Such ultralight scalar fields are motivated  by  string theory
and, in particular, have been studied in the context of the axiverse scenario. For a nonminimally coupled field, the scalar-field energy density behaves as radiation at early times,
which yields a bound  on the coupling constant, $\xi \lesssim 10$, from the primordial nucleosynthesis theory. The first-order perturbations
of the nonminimally coupled field with adiabatic initial conditions cause the gravitational potential to decay on large scales. A comparison of 
the cosmological  data with the theoretical matter power spectrum  yields  the following constraint on the coupling constant: $\xi \lesssim 0.01$. We also consider isocurvature modes in our analysis. We argue that a mix of adiabatic 
and isocurvature initial conditions  for a nonminimally coupled scalar field might allow one to obtain  the usual adiabatic  CDM power spectrum. 

\end{abstract}

\maketitle

\section{Introduction}

A multitude of evidence from astrophysical and cosmological observations supports the existence of dark matter. Such evidence includes galaxy rotation curves \cite{Begeman1991}, large-scale galaxy clustering \cite{SDSS_1, SDSS_2, SDSS_4, SDSS_5, SDSS_14}, cosmological weak gravitational lensing \cite{2007MNRAS.381..702B,Bartelmann:1999yn}, high-redshift supernova 1a \cite{conley2010supernova},
and cosmic microwave background anisotropies \cite{hinshaw:2012aka,Planck2020_1, Planck2020_5, Planck2020_6, Planck2020_8,Planck2020_10,Sievers:2013ica}.  Cold dark matter (CDM) provides a dominant contribution to the dark matter in 
the concordance $\Lambda$CDM  model (e.g. \cite{Dodelson2003} and references therein). However, even after decades of   laboratory and astronomical searches, the nature of cold dark matter has yet to be directly determined.

One of the current leading candidates of cold dark matter is the weakly interacting massive particle (WIMP), which is partly motivated by
the so-called WIMP miracle \cite{Craig:2015xla}. The WIMP miracle  refers to the coincidence that many supersymmetric extensions of the standard particle physics  model  predict the correct cold dark matter abundance for the self-annihilation cross section $\langle \sigma v\rangle \sim 3 \times 10^{-26}\rm{cm^3s^{-1}}$ and masses in the range 100--1000 GeV.The WIMP miracle led to an increased interest in the search of dark matter particles through terrestrial particle physics experiments \cite{Angloher2012, Xenon100_1, Xenon100_2, CDMS-II, LUX, PAMELA_1, PAMELA_2, Adriani2009, PAMELA_3, Fermi_1, Fermi_2, Fermi_3, HEAT, AMS-01, Goodman2011, Fox2012}. However, despite orders-of-magnitude improvement in the
sensitivity of  these experiments,  they have not yet succeeded.

While CMB anisotropy and galaxy clustering data show  the  CDM to be a viable candidate of dark matter for large scales $k < 0.1 \ \rm Mpc^{-1}$,  there could be  multiple issues  with the model at smaller scales. N-body simulations based on the CDM model  overpredict  the number of satellite galaxies of the Milky Way by over an order of magnitude \citep{moore1999dark,klypin1999missing,peebles2010nearby,diemand2007formation,McGaugh2021, Muller2020}. Many  other results also  suggest that the WIMP picture of cold  dark matter predicts more power at galactic scales than has been inferred from observations/N-body simulations  \cite{Perivolaropoulos2021}, e.g.  the cusp-core problem \cite{DeBlok2010} and the
``too big to fail'' issue \cite{GarrisonKimmel2014vqa,BoylanKolchin:2011de}. 
All these issues  have motivated particle physicists and cosmologists   to look  beyond the standard CDM  paradigm  and to  consider alternatives which modify  the CDM model  on galactic scales but  reproduce its success on cosmological scales.

In this paper, we study   ultralight axions (ULA) as a potential
candidate for cold  dark matter.  These ultralight scalar fields
arise naturally in string theory \citep{2006JHEP...06..051S,2007RvMP...79..733D} and become interesting for cosmological observables  when
their masses lie in the range $10^{-33} < m_{\phi} < 10^{-19} \, \rm eV$.  In
particular, these fields have found extensive applications within the framework of the axiverse \citep{arvanitaki2010string,2012PhRvD..86b3508M}. Various studies have shown that the ULA are viable candidates of
cold dark matter on large scales at late times \cite{frieman1995cosmology,coble1997dynamical,hu2000fuzzy,marsh2010ultralight,Marsh2016, UrenaLopez2016, Hlozek2015}. At smaller scales, the ULA behave like an effective
fluid with a  scale-dependent sound speed that  approaches  the speed of light at very small scales \cite{Marsh2016,Hlozek2015}. This implies the suppression of  matter power at small scales, leading to a potential  solution to the small-scale problems.  The
mass range of interest from this perspective is 
$m_{\phi} \gtrsim 10^{-24} \, \rm eV$, and  these models have been studied extensively for various astrophysical and cosmological applications \cite{frieman1995cosmology,coble1997dynamical,hu2000fuzzy,marsh2010ultralight,park2012axion,kobayashi2017lyman, 2017PhRvL.119c1302I,2017MNRAS.471.4606A,2018ApJ...863...73Z,Hui:2016ltb,2016JCAP...04..012S,2017JCAP...07..012S,Hlozek2015,rogers2020strong,2018PhRvD..98h3027B,2019PhRvD..99j3020B,bar2021galactic}. 

The main focus of all the studies involving ULA as  potential CDM candidates has been the minimally coupled scalar fields. It is conceivable that the
gravity sector is more complicated with additional couplings (e.g. \cite{2007PhRvD..75f4020B}). The resulting impact of the more complex dynamics that occur in such 
models  has been studied for  inflationary cosmology,  the growth of perturbations at later times, and the dark energy models close to the current epoch (e.g. \cite{2007PhRvD..75f4020B,2012PhRvD..86b3508M,2008CQGra..25p5001A,2015MNRAS.454.2722H,2015IJMPD..2441002C}).  
In the current study, we extend the analysis to   nonminimally coupled scalar fields.  In these models, the
scalar field has an additional coupling to the gravity sector  (for a more recent application of this additional coupling for
cosmological observables, see \cite{2021arXiv210611971J}; for details of the relevant formulation in gauge-invariant  theory, see \cite{KodamaSasaki1984}). This results
in more complex evolution of both the background and perturbed components and
allows one to consider a wider range of initial conditions. Our main aim is
to compute the matter power spectrum in this case for adiabatic and isocurvature
initial conditions. We also compare our findings against the previous
studies and the observed power spectrum.

This paper has been organized as follows. In Sec. \ref{sec:nonmincoup}, we discuss
in detail the mathematical formulation needed to study  the dynamics of nonminimally coupled scalar fields. In particular, we explicitly  derive the relevant
equations   to study the background and the  perturbed equations and describe
the 
variables employed  to make the problem more tractable. In Sec. \ref{sec:Result}, we present our main results.
In Sec. \ref{sec:concl}, we summarize our results and outline future
perspectives. In  Appendix~\ref{sec:eineq}, we
list the Einstein's equations along with the dynamical equations for the
other components. In Appendix~\ref{sec:incon}, we derive the necessary
initial conditions and discuss salient aspects of the  numerical implementation needed to solve the coupled Einstein-Boltzmann equations in Appendix~\ref{sec:numimpl}. Throughout the paper, we assume a spatially flat universe
and the  best-fit Planck parameters corresponding to it \cite{Planck2020_6}.

\section{Nonminimally coupled real scalar field \label{sec:nonmincoup}}

In this paper, we solve the multiple component system comprised of photons,
massless neutrinos, baryons, cosmological constant, and a nonminimally coupled real scalar field
along with Einstein's equations. In this section,  we describe the scalar-field dynamics
in detail. The relevant equations for other components along with the initial
conditions are given in the Appendixes \ref{sec:eineq} and \ref{sec:incon}. For our work we employ the Newtonian
gauge (e.g. \cite{Bardeen1980, KodamaSasaki1984, Ma1995, Dodelson2003}). In this gauge, the perturbed FRW line element  for a spatially flat universe  is given by:
\begin{equation}
  {\rm d}s^2 = -a^2(\eta)(1 + 2\Psi){\rm d}\eta^2 + a^2(\eta)(1 + 2\Phi)\delta_{ij}{\rm d}x^i{\rm d}x^j.
  \label{eq:pert_met}
\end{equation}
Here $\eta = \int dt/a$ is the conformal time. The functions $\Psi(\eta,\ {\bf x})$ and $\Phi(\eta,\ {\bf x})$ fully specify the  scalar metric perturbations
at first order.

The Lagrangian for the nonminimally coupled scalar field, $\phi$,  is chosen to be (e.g. Kodama and Sasaki \cite{KodamaSasaki1984})
\begin{equation}
    \mathcal{L}_{\phi} = -\frac{1}{2}\sqrt{-g}\big[g^{\mu\nu}\partial_{\mu}\phi\partial_{\nu}\phi + m_{\phi}^2\phi^2 + \xi R\phi^2\big]. \label{eq:Lagrangian}
\end{equation}
Here, $g_{\mu\nu}$ is the metric [Eq.~(\ref{eq:pert_met})], $g$ its determinant, and $m_{\phi}$ is the axion mass. The last term represents the coupling of the scalar field  with the Ricci scalar $R$, with $\xi$ being the dimensionless  coupling constant. The scalar field is minimally coupled when $\xi = 0$. The scalar field is not coupled to any other component. 

By varying the Lagrangian with the field $\phi$, we obtain the field equation given by
\begin{equation}
    \Box{\phi} - \xi R\phi - m_{\phi}^2\phi = 0, \label{eq:EOM}
\end{equation}
where $\Box$ is the d'Alembert operator. 
We can also compute the scalar-field  energy-momentum tensor from the Lagrangian by varying it with respect to the metric $g_{\mu\nu}$. The energy-momentum tensor is computed to be \cite{KodamaSasaki1984}
\begin{eqnarray}
    T^{\mu}_{(\phi)\nu} = \partial^{\mu}\phi\partial_{\nu}\phi - \frac{1}{2}{\delta^{\mu}}_{\nu}\big(\partial^{\lambda}\phi\partial_{\lambda}\phi + m_{\phi}^2\phi^2\big) + \nonumber \\ \xi\big[{G^{\mu}}_{\nu}\phi^2 - {{(\phi^2)_{;}}^{\mu}}_{\nu} + {\delta^{\mu}}_{\nu}\Box(\phi^2)\big]. \label{eq:EM_tensor}
\end{eqnarray}
Here ${G^{\mu}}_{\nu}$ is the Einstein tensor, and the  semicolon represents the covariant derivative.
In this paper, we solve the coupled  system  up to first order in perturbation theory. Therefore, the scalar field can be decomposed into a homogeneous and an inhomogeneous component, $\phi(\eta,\ {\bf x}) = \phi_0(\eta) + \delta\phi(\eta,\ {\bf x})$, and all the terms that are second order in $\delta\phi(\eta,\ {\bf x})$ are dropped.

\subsection{Background equations \label{sec:Background_Equations}}

The zeroth order equation of motion for the scalar field can be
obtained from Eq.~(\ref{eq:EOM}):
\begin{equation}
    \Ddot{\phi}_0 + 2aH\Dot{\phi}_0 + a^2m_{\phi}^2\phi_0 + 3\xi a^2H^2(1-3w)\phi_0 = 0. \label{eq:bg_KG_eqn}
\end{equation}
The overdot represents a derivative with respect to $\eta$. Note that $H$ is the Hubble parameter defined as $H = \Dot{a}/a^2$, and $w$ is the equation of state of the universe defined as the ratio of the total pressure to the total energy density of all the components constituting the universe:
\begin{equation}
  w = \frac{\sum_i p_i}{\sum_i \rho_i}.
  \label{eq:eqosta}
\end{equation}

The scalar-field energy density can be  computed using Eq.~(\ref{eq:EM_tensor}):
\begin{equation}
  \rho_{\phi} = \frac{1}{2}\frac{\Dot{\phi_0}^2}{a^2} + \frac{1}{2}m_{\phi}^2\phi_0^2 + 3\xi H^2\bigg[\phi_0^2 + \frac{2}{aH}\phi_0\Dot{\phi_0}\bigg].
  \label{eq:backeneden}
\end{equation}
Equations~(\ref{eq:bg_KG_eqn})--(\ref{eq:backeneden}),~(\ref{eq:dens_parameter_evolution}),~(\ref{eq:Einstein_1}),
  and~(\ref{eq:Einstein_2})  describe the background evolution
  of  all the relevant variables. This system of equations could be stiff even for the minimally coupled case. To overcome this issue,  we define a new set of variables---$\widetilde{\Omega}_{\phi}$ and $\theta$; the choice of these variables is
motivated by similar variables used by Ure{\~n}a-L{\'o}pez and Gonzalez-Morales \cite{UrenaLopez2016}:
\begin{eqnarray}
    \sqrt{\widetilde{\Omega}_{\phi}}\sin\bigg(\frac{\theta}{2}\bigg) = \sqrt{\frac{4\pi G}{3}}\frac{\Dot{\phi}_0}{aH}, \label{eq:transformation1} \\
    \sqrt{\widetilde{\Omega}_{\phi}}\cos\bigg(\frac{\theta}{2}\bigg) = -\sqrt{\frac{4\pi G}{3}}\frac{m_{\phi}\phi_0}{H}. \label{eq:transformation2}
\end{eqnarray}
Here $G$ is the gravitational constant. The background equation of motion, Eq.~(\ref{eq:bg_KG_eqn}), in terms of the  new variables is given by
\begin{eqnarray}
    \widetilde{\Omega}_{\phi}' = 3(w + \cos\theta)\widetilde{\Omega}_{\phi} + 6\frac{\xi}{y}(1 - 3w)\widetilde{\Omega}_{\phi}\sin\theta, \label{eq:Omega_phi_prime} \\
    \theta' = -3\sin\theta + y + 6\frac{\xi}{y}(1-3w)(1 + \cos\theta). \label{eq:theta_prime}
\end{eqnarray}
Here the  prime denotes a derivative with respect to $\ln{a}$ and  $y = 2m_{\phi}/H$. Since $a$ is a monotonically increasing function of $\eta$, we can use $\ln{a}$ as a time variable instead of $\eta$.

The scalar-field  energy density $\rho_{\phi}$ and the equation of state of the
scalar field $w_{\phi}$ can be expressed in terms of these new variables as
\begin{equation}
    \Omega_{\phi} = \frac{8\pi G \rho_\phi}{3H^2} = \widetilde{\Omega}_{\phi}\bigg[1 + 12\xi\bigg(\frac{1 + \cos\theta}{y^2} - \frac{\sin\theta}{y}\bigg)\bigg], \label{eq:Omega_a_defn}
\end{equation}
\begin{multline}
    \bigg[1 + 12\frac{\xi}{y^2}(1+\cos\theta)\{1 - (1-6\xi)\Omega_{\phi}\} - 12\frac{\xi}{y}\sin\theta\bigg]w_{\phi} \\ = -(1-4\xi)\cos\theta + 12\frac{\xi}{y^2}(1-6\xi)(1+\cos\theta)\bigg(\frac{\Omega_{\gamma} + \Omega_{\nu}}{3} - \Omega_{\Lambda}\bigg) \\ + 24\frac{\xi^2}{y^2}(1 + \cos\theta). \label{eq:w_phi_defn}
\end{multline}
Note that $\Omega_{\gamma}$, $\Omega_{\nu}$, and $\Omega_{\Lambda}$ are the photon, neutrino and dark energy density parameters, respectively.  Note that for the minimally coupled scalar field  ($\xi = 0$), these equations can be simplified to $\widetilde{\Omega}_{\phi} = \Omega_{\phi}$ and $w_{\phi} = -\cos\theta$, respectively (for details, see Ure{\~n}a-L{\'o}pez and Gonzalez-Morales \cite{UrenaLopez2016}).

By numerically solving Eqs.(\ref{eq:Omega_phi_prime}) and~(\ref{eq:theta_prime})
along with the background Einstein equations and the equations of motion for other components in the universe given in Appendix \ref{sec:Appendix_Background_Equations}, we obtain the
evolution of all the relevant background variables.

\subsection{First order equations \label{sec:First_Order_Equations}}

Using the perturbed metric [Eq.~(\ref{eq:pert_met})], Eq.~(\ref{eq:EOM}) yields  the following equation of motion for the scalar field  in the  Fourier space at first order:
\begin{multline}
    \delta\Ddot{\phi} + 2aH\delta\Dot{\phi} + \big[k^2 + a^2m_{\phi}^2 + 3\xi a^2H^2(1 - 3w)\big]\delta\phi = \\ \big(\Dot{\Psi} - 3\Dot{\Phi}\big)\Dot{\phi}_0 - a^2\big[2m_{\phi}^2\Psi + \xi\big\{\delta R + 6H^2(1-3w)\Psi\big\}\big]\phi_0. \label{eq:pert_KG_eqn}
\end{multline}
Here, $k$ is the wave number of the mode, and $\delta R$ is the perturbed part of the  Ricci scalar.

To solve Eq.~(\ref{eq:pert_KG_eqn}), which could be stiff,  we again define two new variables, $\alpha$ and $\beta$, which are motivated by similar variables used by  Ure{\~n}a-L{\'o}pez and Gonzalez-Morales \cite{UrenaLopez2016}:
\begin{eqnarray}
    \sqrt{\widetilde{\Omega}_{\phi}}\alpha\cos\bigg(\frac{\theta - \beta}{2}\bigg) = -\sqrt{\frac{16\pi G}{3}}\frac{\delta\Dot{\phi}}{aH}, \\
    \sqrt{\widetilde{\Omega}_{\phi}}\alpha\sin\bigg(\frac{\theta - \beta}{2}\bigg) = -\sqrt{\frac{16\pi G}{3}}\frac{m_{\phi}\delta\phi}{H}.
\end{eqnarray}
We can express Eq.~(\ref{eq:pert_KG_eqn}) in terms of the new variables:
\begin{multline}
    \alpha' = -\frac{3}{2}\alpha[\cos\theta + \cos(\theta - \beta)] - \frac{k^2}{2k_J^2}\alpha\sin(\theta - \beta) \\ - (\Psi' - 3\Phi')\bigg[\sin\bigg(\theta - \frac{\beta}{2}\bigg) + \sin\bigg(\frac{\beta}{2}\bigg)\bigg] \\ - y\Psi\bigg[\cos\bigg(\theta - \frac{\beta}{2}\bigg) + \cos\bigg(\frac{\beta}{2}\bigg)\bigg] \\ -\xi\bigg[\frac{3\alpha}{y}(1-3w)\{\sin\theta + \sin(\theta - \beta)\} \\ + \frac{2}{y}\bigg(\frac{\delta R}{H^2} + 6(1-3w)\Psi\bigg)\bigg\{\cos\bigg(\theta - \frac{\beta}{2}\bigg) + \cos\bigg(\frac{\beta}{2}\bigg)\bigg\}\bigg], \label{eq:alpha_eqn}
\end{multline}
\begin{multline}
    \beta' = -3[\sin\theta + \sin(\theta - \beta)] - \frac{k^2}{k_J^2}[1-\cos(\theta-\beta)] \\ - 2\bigg(\frac{\Psi' - 3\Phi'}{\alpha}\bigg)\bigg[\cos\bigg(\frac{\beta}{2}\bigg) - \cos\bigg(\theta - \frac{\beta}{2}\bigg)\bigg] \\ - 2\frac{y\Psi}{\alpha}\bigg[\sin\bigg(\theta - \frac{\beta}{2}\bigg) - \sin\bigg(\frac{\beta}{2}\bigg)\bigg] \\ + \frac{2\xi}{\alpha}\bigg[\frac{3\alpha}{y}(1-3w)\{\cos\theta + \cos(\theta - \beta)\} \\ - \frac{2}{y}\bigg(\frac{\delta R}{H^2} + 6(1 - 3w)\Psi\bigg)\bigg\{\sin\bigg(\theta - \frac{\beta}{2}\bigg) - \sin\bigg(\frac{\beta}{2}\bigg)\bigg\}\bigg]. \label{eq:beta_eqn}
\end{multline}
Here $k_J = a\sqrt{m_{\phi}H}$; $k_J$ acts as an effective Jeans' scale in the
dynamics of a perturbed scalar field. In the minimally coupled case,
perturbations corresponding to scales $k > k_J$ cannot grow. The definition of
$k_J$ adequately captures the dependence of this scale on time and the mass
of the scalar field even in the more general case we consider here.

From the energy-momentum tensor, Eq.~(\ref{eq:EM_tensor}), we can define four fluid quantities at first order. These quantities are the  density perturbation $\delta_{\phi}$,  the irrotational  component of the bulk velocity  $v_{\phi}$, the isotropic pressure perturbation $\pi_{\phi}$ and the anisotropic  stress  $\Pi_{\phi}$. These quantities can be expressed in terms of the new variables as follows:
\begin{multline}
    \bigg[1 + 12\frac{\xi}{y}\bigg(\frac{1 + \cos\theta}{y} - \sin\theta\bigg)\bigg]\delta_{\phi} = - \Psi(1 - \cos\theta) - \\ \alpha\sin\bigg(\frac{\beta}{2}\bigg) + 12\frac{\xi}{y}\bigg[2\bigg(\frac{1 + \cos\theta}{y}\bigg)\bigg(\Phi' - \Psi + \frac{k^2}{3a^2H^2}\Phi\bigg) \\ - (\Phi' - 2\Psi)\sin\theta + \frac{\alpha}{y}\bigg(1 + \frac{k
        ^2}{3a^2H^2}\bigg)\bigg\{\sin\bigg(\theta - \frac{\beta}{2}\bigg) - \\ \sin\bigg(\frac{\beta}{2}\bigg)\bigg\} + \alpha\cos\bigg(\theta - \frac{\beta}{2}\bigg)\bigg]
    \label{eq:denper_def}
\end{multline}

\begin{multline}
    \bigg[1 + 12\frac{\xi}{y}\bigg(\frac{1 + \cos\theta}{y} - \sin\theta\bigg)\bigg](1 + w_{\phi})v_{\phi} = \\ \frac{ak}{H}\frac{\alpha}{y}\bigg[\cos\bigg(\theta - \frac{\beta}{2}\bigg) - \cos\bigg(\frac{\beta}{2}\bigg)\bigg] - 4\frac{k}{aH}\frac{\xi}{y}\bigg[\Psi\sin\theta \\ + \alpha\cos\bigg(\theta - \frac{\beta}{2}\bigg) + \frac{2}{y}(\Phi' - \Psi)(1 + \cos\theta) - \\ \frac{\alpha}{y}\bigg\{\sin\bigg(\theta - \frac{\beta}{2}\bigg) - \sin\bigg(\frac{\beta}{2}\bigg)\bigg\}\bigg]
\end{multline}

\begin{multline}
  \bigg[1 + 12\frac{\xi}{y}\bigg(\frac{1 + \cos\theta}{y} - \sin\theta\bigg)\bigg]w_{\phi}\pi_{\phi} = -\Psi(1-\cos\theta) \\ - \alpha\sin\bigg(\theta - \frac{\beta}{2}\bigg) + 2\xi\bigg[2\Psi(1 - \cos\theta) + 2\alpha\sin\bigg(\theta - \frac{\beta}{2}\bigg) - \\ \frac{9}{y^2}(1 + \cos\theta)\bigg(\Phi'' + 3\Phi' - \Psi' - 3\Psi - \frac{3}{2}(1 + w)(\Phi' - 2\Psi) + \\ \frac{k^2}{3a^2H^2}(\Phi + \Psi)\bigg) + 6\frac{\alpha}{y^2}\bigg\{\sin\bigg(\theta - \frac{\beta}{2}\bigg) - \sin\bigg(\frac{\beta}{2}\bigg)\bigg\}\times \\ \bigg\{w + 2\xi(1 - 3w) + \frac{k^2}{9a^2H^2}\bigg\} + 2\xi\frac{\delta R}{m_{\phi^2}}(1 + \cos\theta)\bigg]
  \label{eq:prespert}
\end{multline}

\begin{multline}
    \bigg[1 + 12\frac{\xi}{y}\bigg(\frac{1 + \cos\theta}{y} - \sin\theta\bigg)\bigg]w_{\phi}\Pi_{\phi} = -\frac{\xi k^2}{a^2m_{\phi}^2} \times \\ \bigg[(1+\cos\theta)(\Phi + \Psi) + \alpha\bigg\{\sin\bigg(\theta - \frac{\beta}{2}\bigg) - \sin\bigg(\frac{\beta}{2}\bigg)\bigg\}\bigg] \label{eq:axion_anis_stress}
\end{multline}
Equations~(\ref{eq:denper_def})--(\ref{eq:axion_anis_stress}) provide 
source terms to the first order Einstein's equations [Eqs.~(\ref{eq:einfir})--(\ref{eq:Anisotropic_stress_equation})].
We note that the anisotropic stress $\Pi_{\phi}$ is nonzero only for
the nonminimally coupled case. 

The system of relevant equations constitutes  the first order Klein-Gordon equation, Eqs.~(\ref{eq:alpha_eqn}) and (\ref{eq:beta_eqn}),  along with the first order Einstein equations and the Boltzmann equations for other components of the universe given in Appendix \ref{sec:Appendix_First_Order_Equations}. For comparison with the observed matter power spectrum at the current
epoch, the quantity of direct interest is the scalar-field  density perturbation $\delta_{\phi}$ [Eq.~(\ref{eq:denper_def})], which can be determined from the solutions of  $\alpha$ and $\beta$.

\section{Results \label{sec:Result}}

In addition to the zeroth and the first order dynamical equations for the nonminimally coupled scalar field, we need to consider
other components in the universe which either dominate at early times (photons and neutrinos), make a comparable contribution 
(baryons), or  dominate at late times (cosmological constant). For the cosmological constant, we only consider the background evolution
as it does not couple to first order equations (e.g. \cite{Ma1995}).   The coupled zeroth and first order
Boltzmann equations of all these components along with Einstein's equations are   given in Appendix~\ref{sec:eineq}.
These equations are solved along with the scalar-field equations in  Secs.~\ref{sec:Background_Equations} and  \ref{sec:First_Order_Equations}.   In Appendix~\ref{sec:incon},  we derive the relevant initial conditions. For our
work, we consider two
initial conditions: adiabatic and isocurvature.

We first discuss the zeroth order solutions. 

\subsection{Background solution}

\begin{figure}[!ht]
    \centering
    \begin{subfigure}[b]{0.99\linewidth}
        \includegraphics[width = 0.99\linewidth]{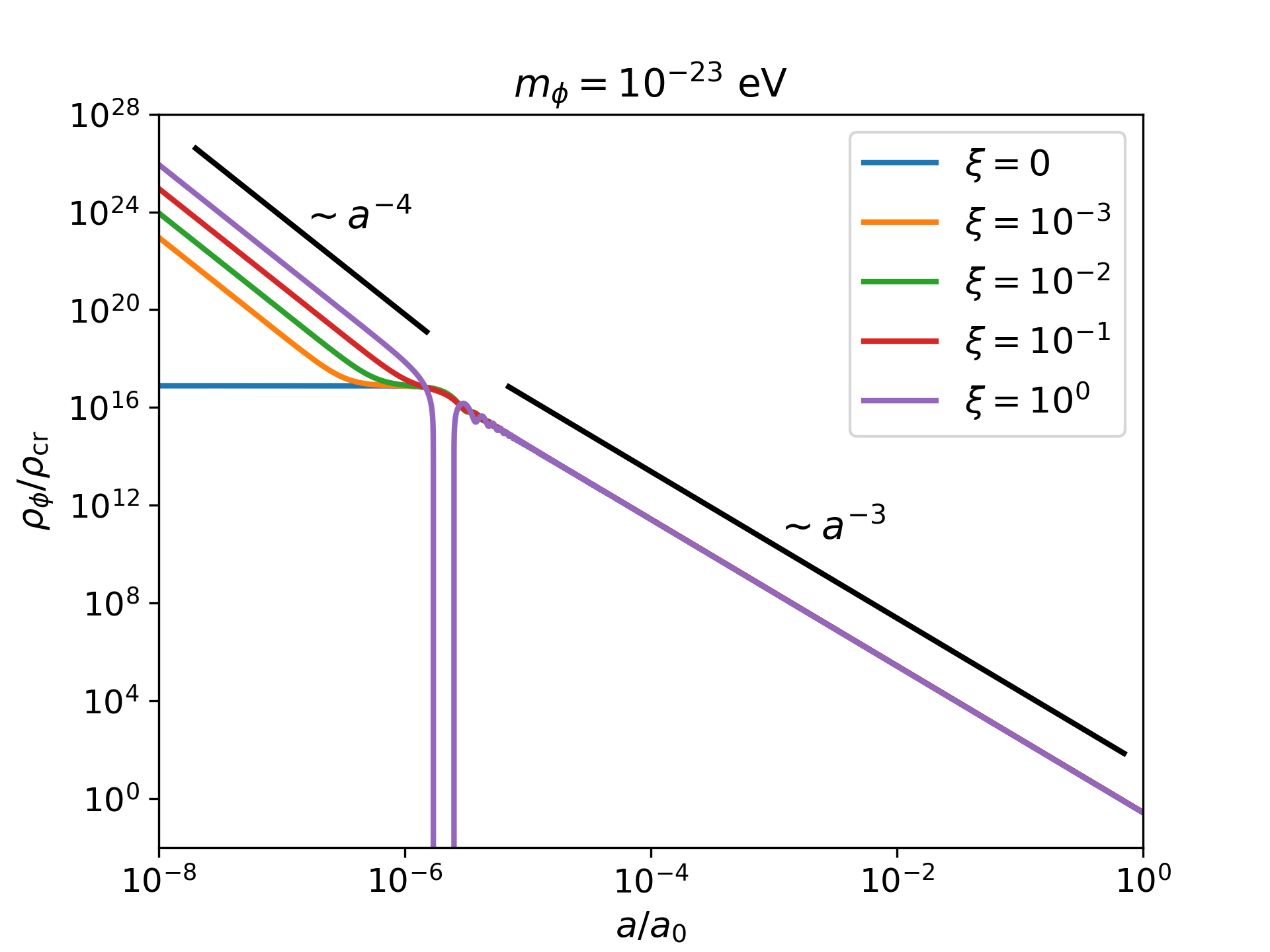}
        \caption{Energy density as a function of scale factor for different values of $\xi$.}
        \label{fig:bg_1}
    \end{subfigure}
    
    \begin{subfigure}[b]{0.99\linewidth}
        \includegraphics[width = 0.99\linewidth]{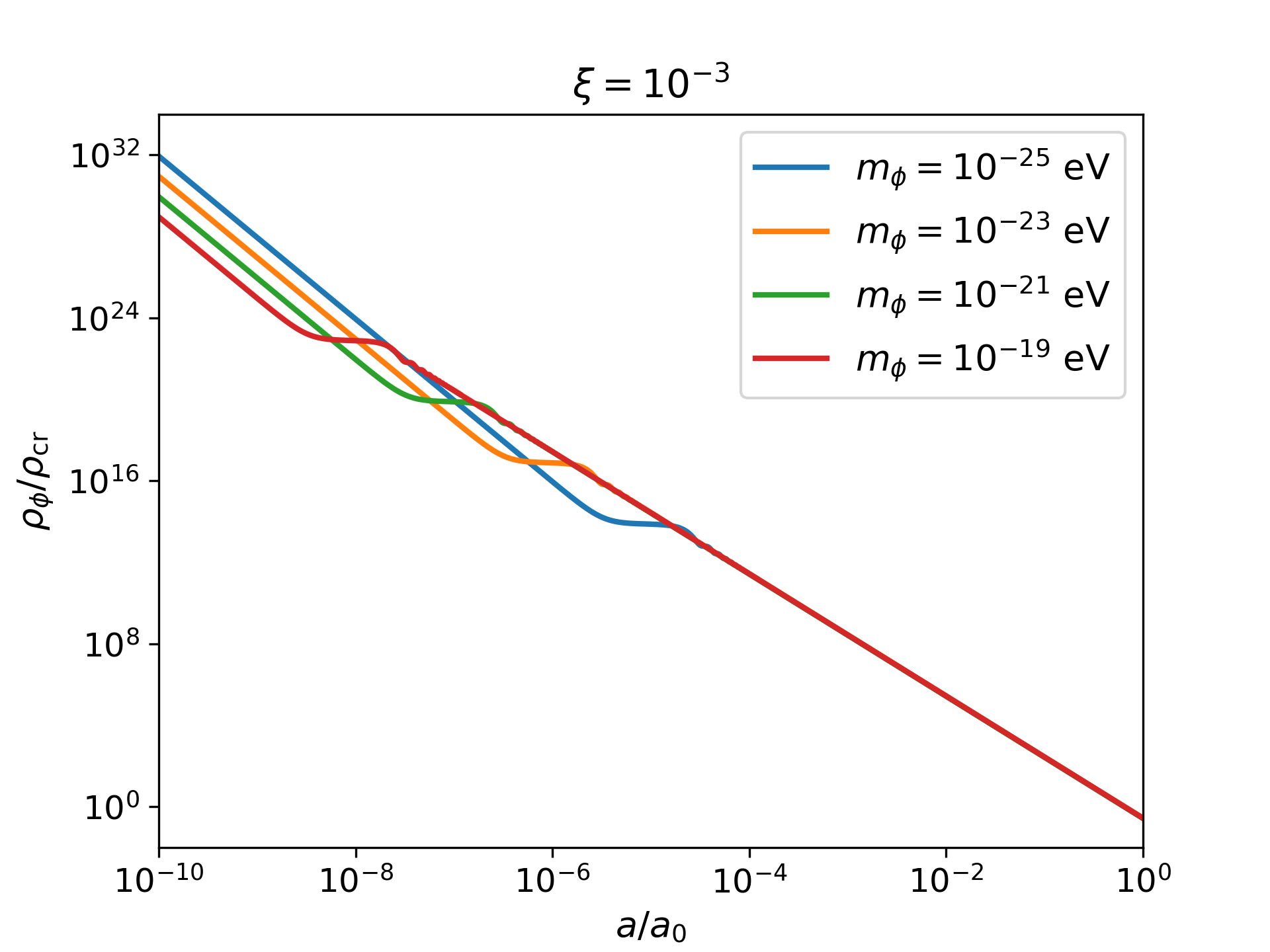}
        \caption{Energy density as a function of scale factor for different values of $m_{\phi}$.}
        \label{fig:bg_2}
    \end{subfigure}
    
    \caption{Evolution of ULA energy density displayed in terms of the critical density at the current epoch: $\rho_{\rm cr} = 3H^2(t_0)/8\pi G$}
    \label{fig:background}
\end{figure}

Figures~\ref{fig:bg_1} and~\ref{fig:bg_2}  display the evolution of the energy density of the scalar field  for different values of $\xi$ and scalar-field  masses. First, we briefly summarize the well-known
behavior for a minimally coupled field. During the initial phase, $y \ll 1$ and $\theta \ll 1$ (see Appendix \ref{sec:Background_Initial_Conditions} for details). During this phase, the friction term [the term proportional to $\dot\phi_0$ in Eq.~(\ref{eq:bg_KG_eqn})] dominates. This  causes   the scalar-field
energy density [Eq.~(\ref{eq:backeneden})] to be  nearly constant, resulting in  the scalar field behaving as
a cosmological constant. For $m_{\phi} \gtrsim H$, the scalar field starts oscillating around $\phi_0 = 0$. This phase culminates with the oscillation frequency converging to $\simeq m_\phi t$. For $m_\phi t \gg 1$, the rate of this oscillation far exceeds the 
expansion timescale, and as our aim  is to track the evolution of relevant quantities on the expansion timescale, we define time-averaged quantities
where the time average is over rapid oscillations.  During this phase, the time-averaged   $\phi_0$ vanishes  and the  time-averaged    scalar-field energy
density falls as $a^{-3}$ or  it behaves as nonrelativistic matter as cold dark matter.   For numerical stability, we employ the WKB approximation during the oscillatory phase and  switch from
the oscillatory solutions to the time-averaged  evolution at $\theta = \pi/2$ (for further details see e.g. \cite{UrenaLopez2016}).

Figure~\ref{fig:bg_1} shows that for nonzero $\xi$, the scalar-field  energy density scales as $a^{-4}$ during the initial phase like  the energy density of  relativistic components (photons and neutrinos).\footnote{We note that this yields an additional constraint on the nonminimally coupled case from primordial nucleosynthesis, which we discuss later.} In the intermediate phase, for $\xi \lesssim 10^{-1}$, the scalar field makes a transition to a phase in which its energy density is constant. For $\xi \gtrsim 1$, the energy density in the intermediate phase can become negative. At a later stage, $a \simeq 10^{-6}$, the energy density starts falling off as $a^{-3}$ for all values of $\xi$ for $m_{\phi} = 10^{-23}$ eV. Figure~\ref{fig:bg_2} shows the impact  of changing $m_\phi$ for a  fixed  $\xi$ on the energy density. For larger $m_\phi$, the field enters both the constant energy density phase and the $a^{-3}$ phase earlier. 

In this paper, we focus on the range of values of $\xi$ for which the scalar-field  energy density is always a positive quantity. Thus, we work with $\xi \lesssim 1$.

\begin{figure}[!ht]
    \centering
    \includegraphics[width = .99\linewidth]{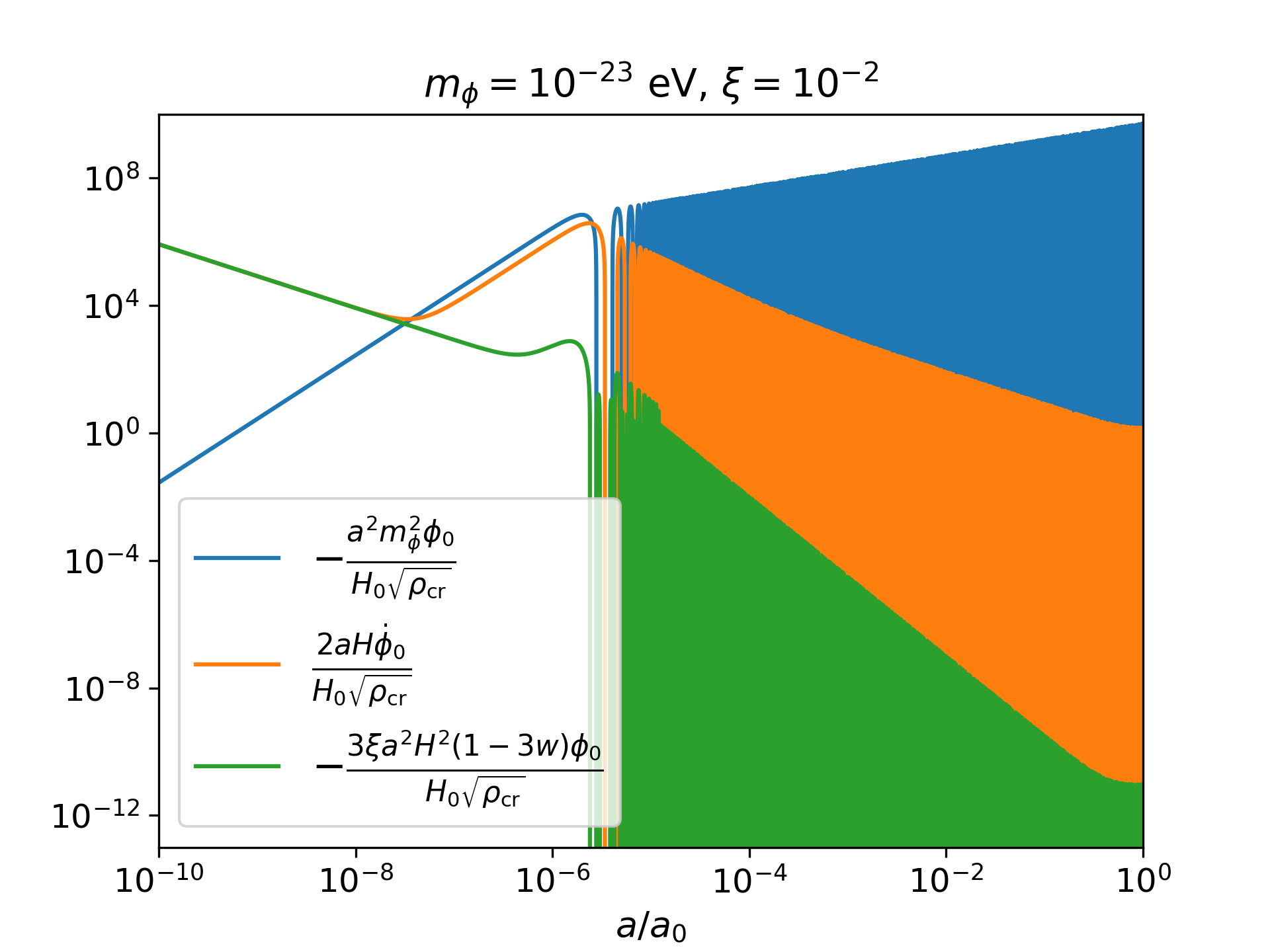}
    \caption{Different terms of Eq.~(\ref{eq:bg_KG_eqn}) shown as a function of time.}
    \label{fig:bg_term_comparison}
\end{figure}

In  Fig.~\ref{fig:bg_term_comparison}, we further investigate the relative contribution of  different terms in  the equation of motion  of nonminimally coupled scalar fields (Eq.~(\ref{eq:bg_KG_eqn})). It is seen that
the term proportional to the nonminimal coupling $\xi$ always dominates at initial times.  This  term is $\propto H^2$ and follows from the corresponding term
in the scalar-field  Lagrangian, Eq.~(\ref{eq:Lagrangian}), which is  proportional to the Ricci scalar $R \propto H^2$. This causes the energy density of the scalar field   to behave as  radiation during  early times. At later times, the nonminimal coupling term  becomes smaller than other terms, mainly owing to the rapid decrease in $H^2$, and  the nonminimally coupled model behaves as a  minimally coupled model. 

This shows that the  main difference in the  background evolution between  the minimally and nonminimally coupled scalar
fields is  during the initial phase:  The energy density of the
minimally coupled scalar field  ($\xi = 0$) is constant during this phase while
 it falls as   $a^{-4}$ for nonzero $\xi$.

\begin{figure}[!ht]
    \centering
    \includegraphics[width = .99\linewidth]{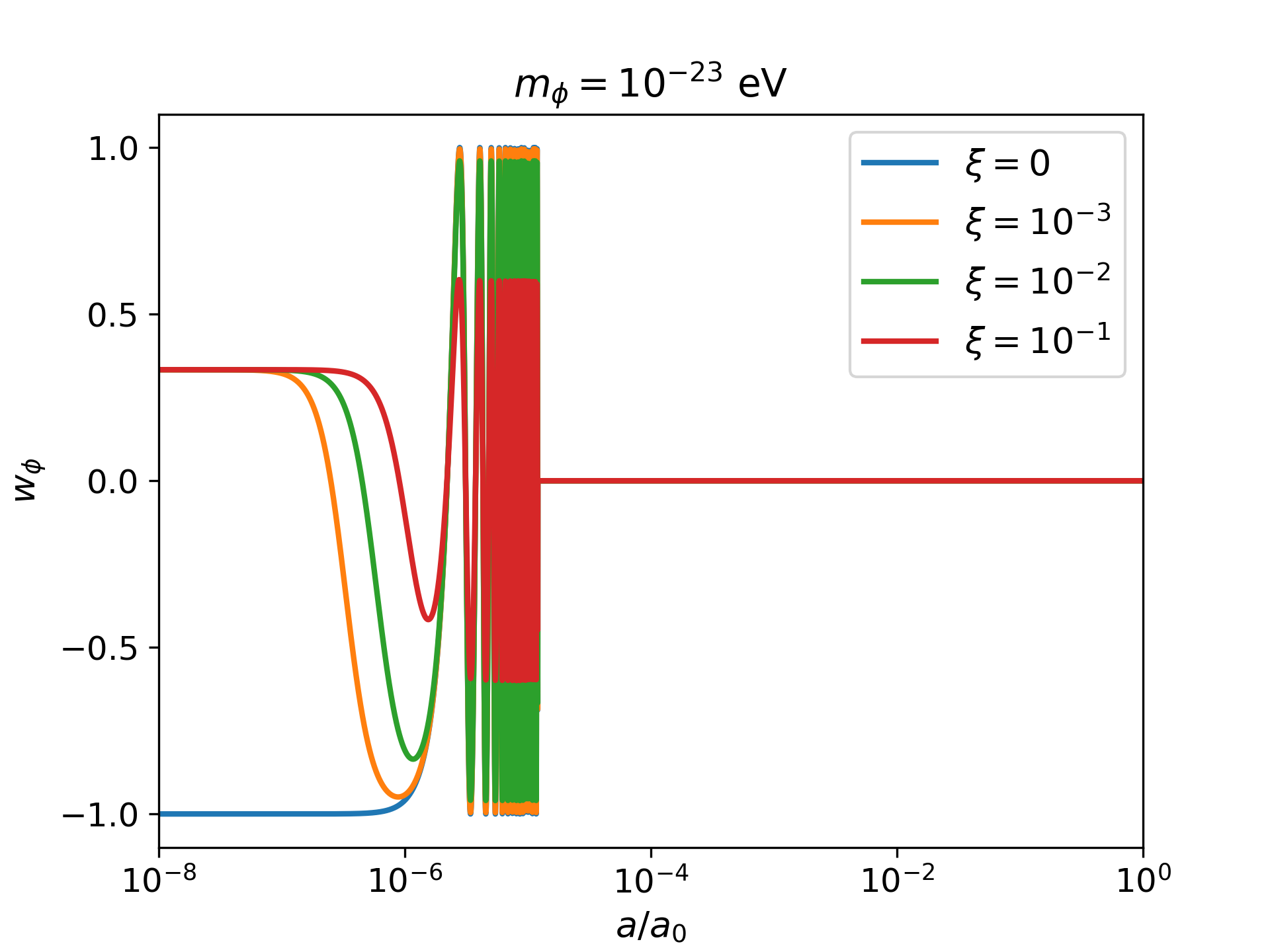}
    \caption{Evolution of the scalar-field  equation of state $w_{\phi}$ displayed   for different values of $\xi$. The plot  is exact for  $a \lesssim 10^{-5}$. For $a \gtrsim 10^{-5}$,  the time average of $w_{\phi}$ is shown (see text for more details).}
    \label{fig:bg_3}
\end{figure}

The evolution of the scalar-field  equation of state $w_{\phi}$ is shown in Fig.~\ref{fig:bg_3}. In the figure, we switch from the full solution to the time-averaged solution at $a=10^{-5}$. We tried to show the oscillations in $w_{\phi}$; for the actual solution, we switch to the time-averaged solution at $\theta = \pi/2$, as discussed above. As follows from the discussion above, the minimally coupled scalar field  starts  as a cosmological constant with $w_{\phi} = -1$ while the  nonminimally coupled scalar field  has the equation of state, $w_{\phi} = 1/3$ (or it mimics radiation) during the initial phase, irrespective of the value of $\xi$. This radiationlike behaviour of a nonminimally coupled scalar field  at early times
can also be inferred from Eq.~(\ref{eq:w_phi_defn}). At early times, $y \ll 1$ and $\theta \ll 1$ (Appendix \ref{sec:Background_Initial_Conditions}). Further, using the fact that   $\Omega_{\gamma}$ and $\Omega_{\nu}$ dominate other  density parameters during the radiation-dominated era, $w_{\phi}$ becomes
\begin{equation}
    w_{\phi} \simeq \frac{8\frac{\xi}{y^2}(1 - 6\xi) + 48\frac{\xi^2}{y^2}}{24\frac{\xi}{y^2}} = \frac{1}{3}. \label{eq:w_phi_approximation}
\end{equation}

During intermediate times, the equation of state $w_{\phi}$ tends to $-1$ for the nonminimally coupled case also. For $m_\phi \gtrsim H$, $w_{\phi}$ oscillates between positive and negative values, averaging to zero, for all the cases.

\subsection{First order solution}

\begin{figure}[!ht]
    \centering
    \begin{subfigure}[b]{0.99\linewidth}
        \includegraphics[width = 0.99\linewidth]{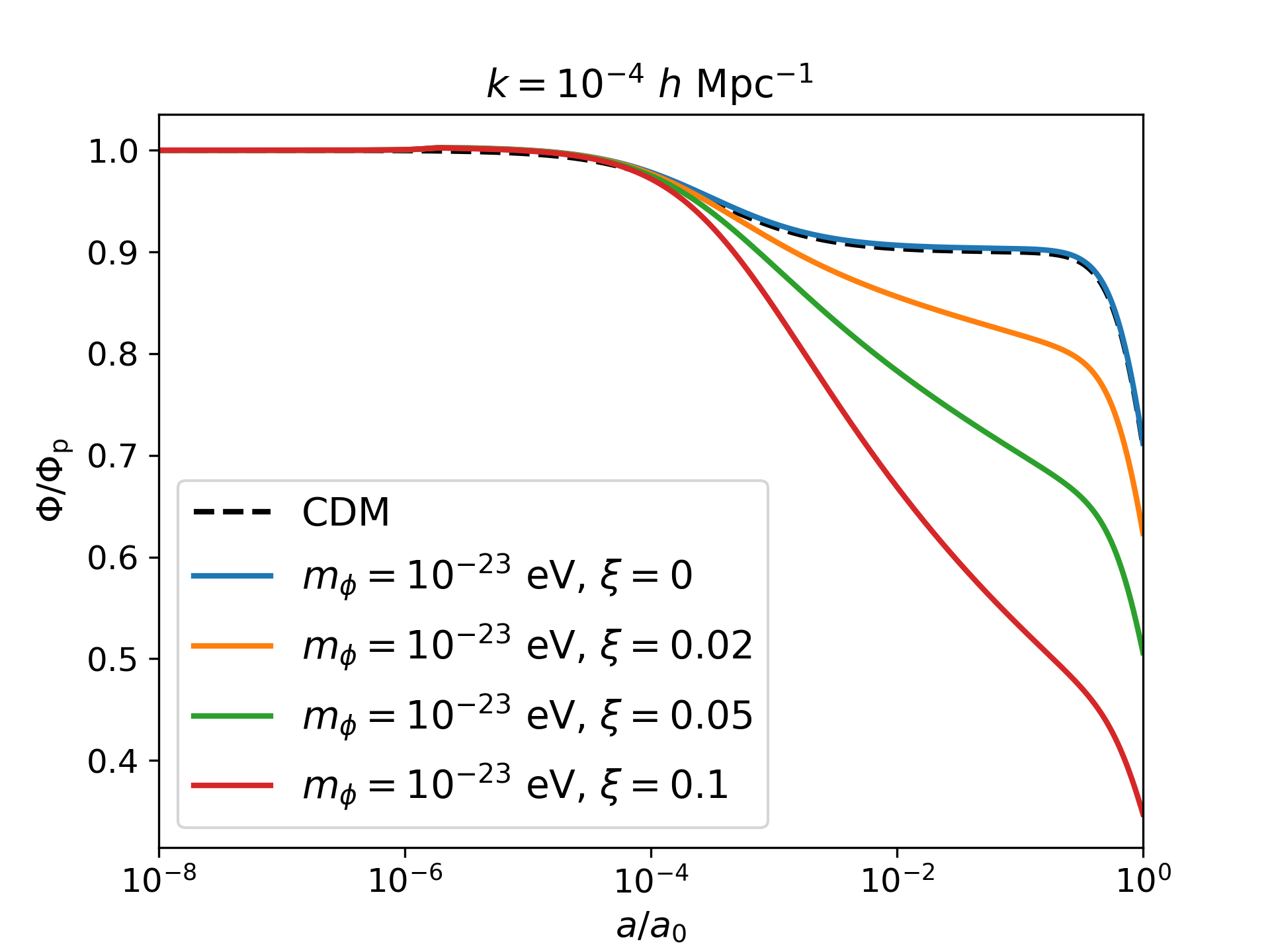}
    \end{subfigure}
    
    \begin{subfigure}[b]{0.99\linewidth}
        \includegraphics[width = 0.99\linewidth]{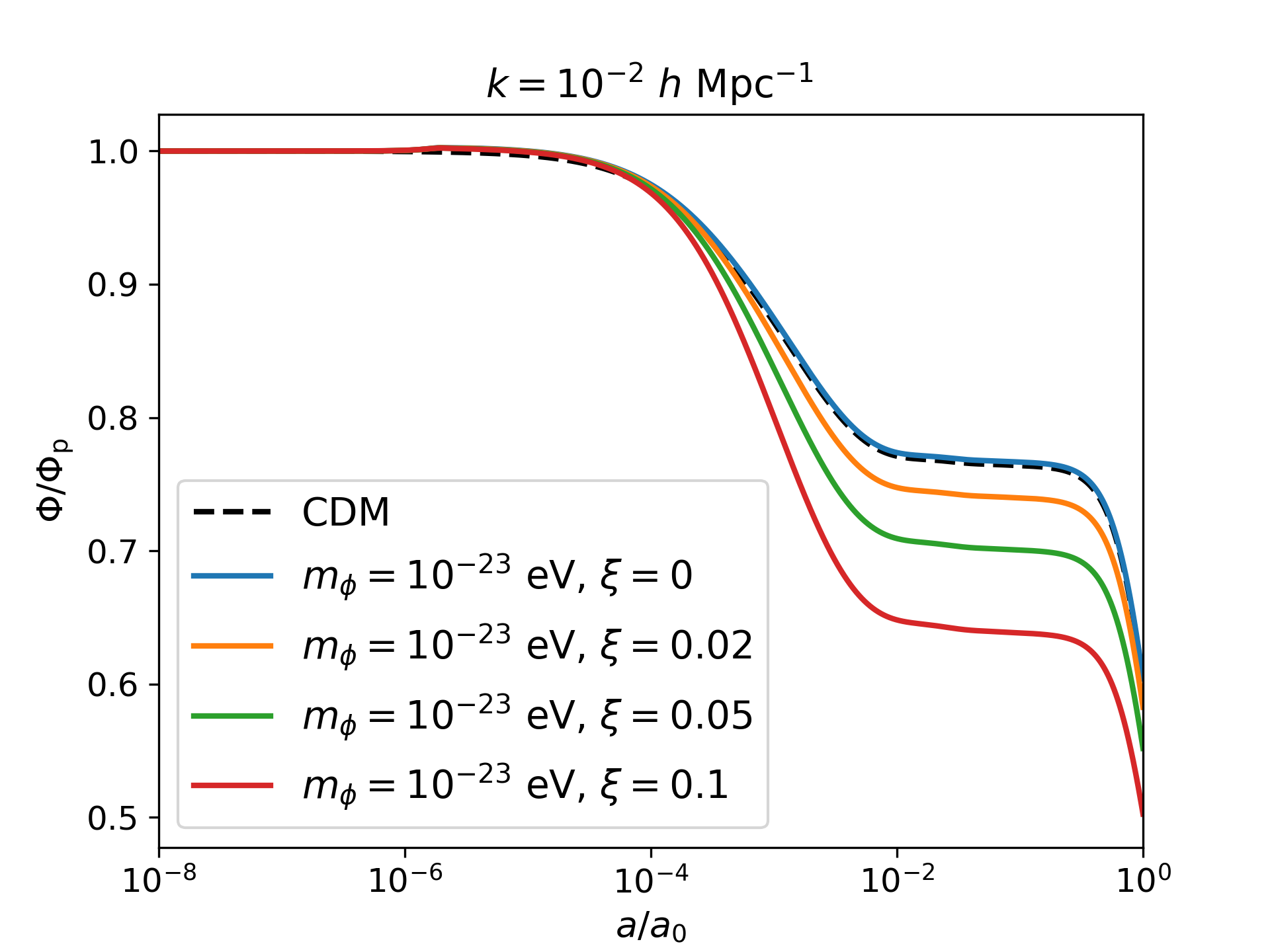}
    \end{subfigure}
    
    \begin{subfigure}[b]{0.99\linewidth}
        \includegraphics[width = 0.99\linewidth]{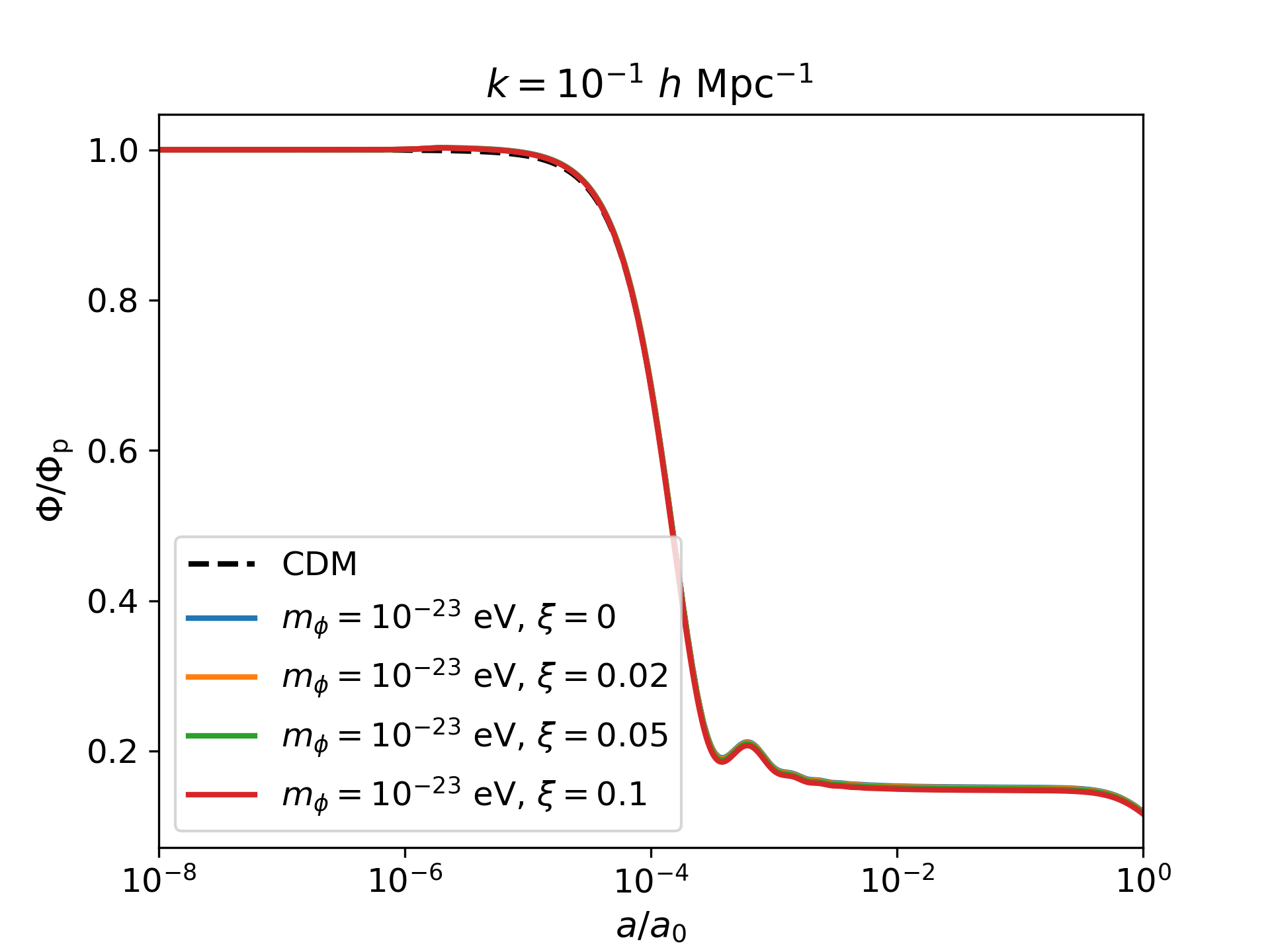}
    \end{subfigure}
    
    \caption{Scalar metric perturbation $\Phi$ (adiabatic mode) shown as a  function of  the scale factor for different values of $k$. Note that $\Phi_p$ is the primordial value of $\Phi$. \label{fig:Phi}}
\end{figure}

\begin{figure}[!ht]
    \centering
    \begin{subfigure}[b]{0.99\linewidth}
        \includegraphics[width = 0.99\linewidth]{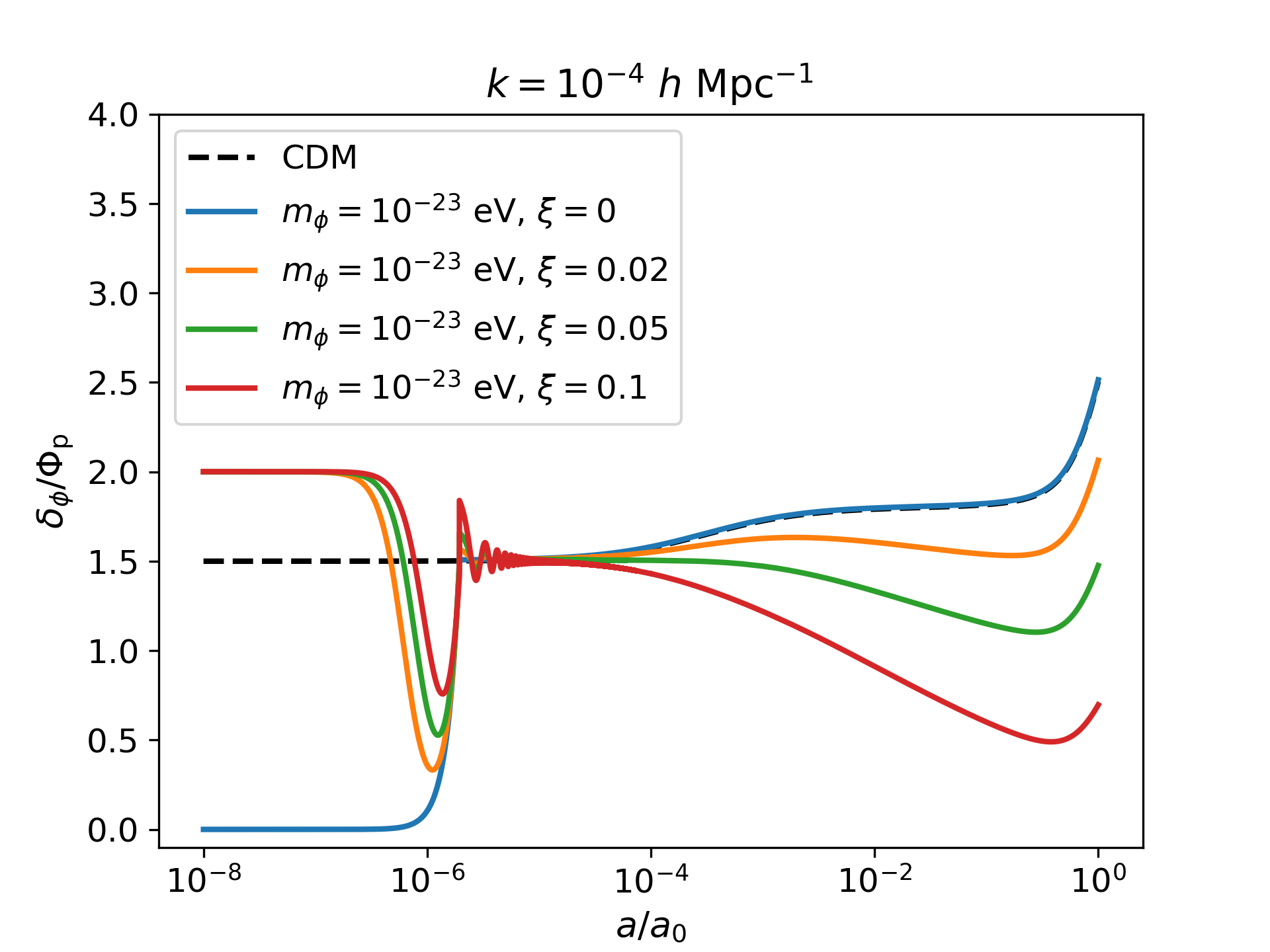}
    \end{subfigure}
    
    \begin{subfigure}[b]{0.99\linewidth}
        \includegraphics[width = 0.99\linewidth]{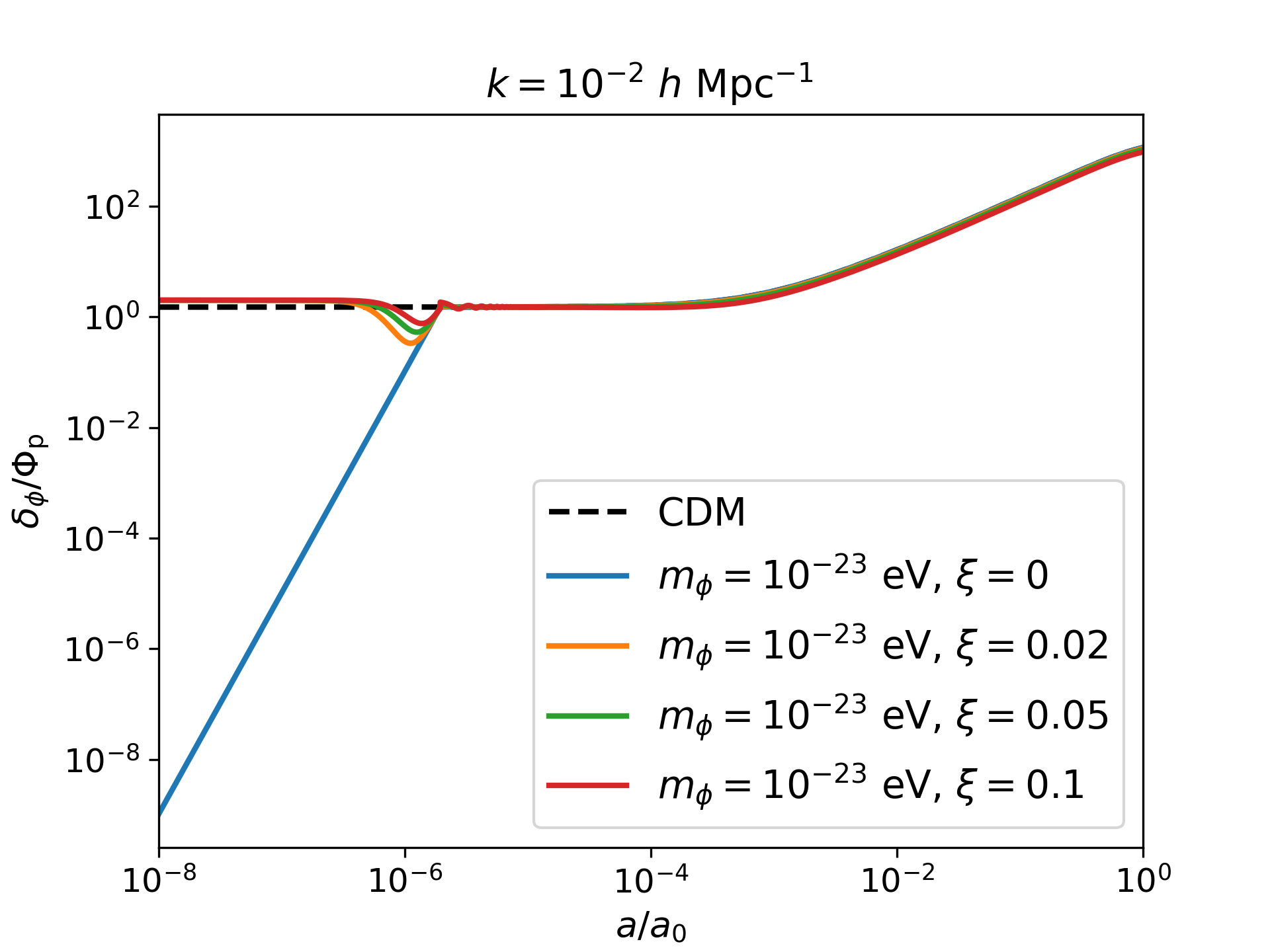}
    \end{subfigure}
    
    \begin{subfigure}[b]{0.99\linewidth}
        \includegraphics[width = 0.99\linewidth]{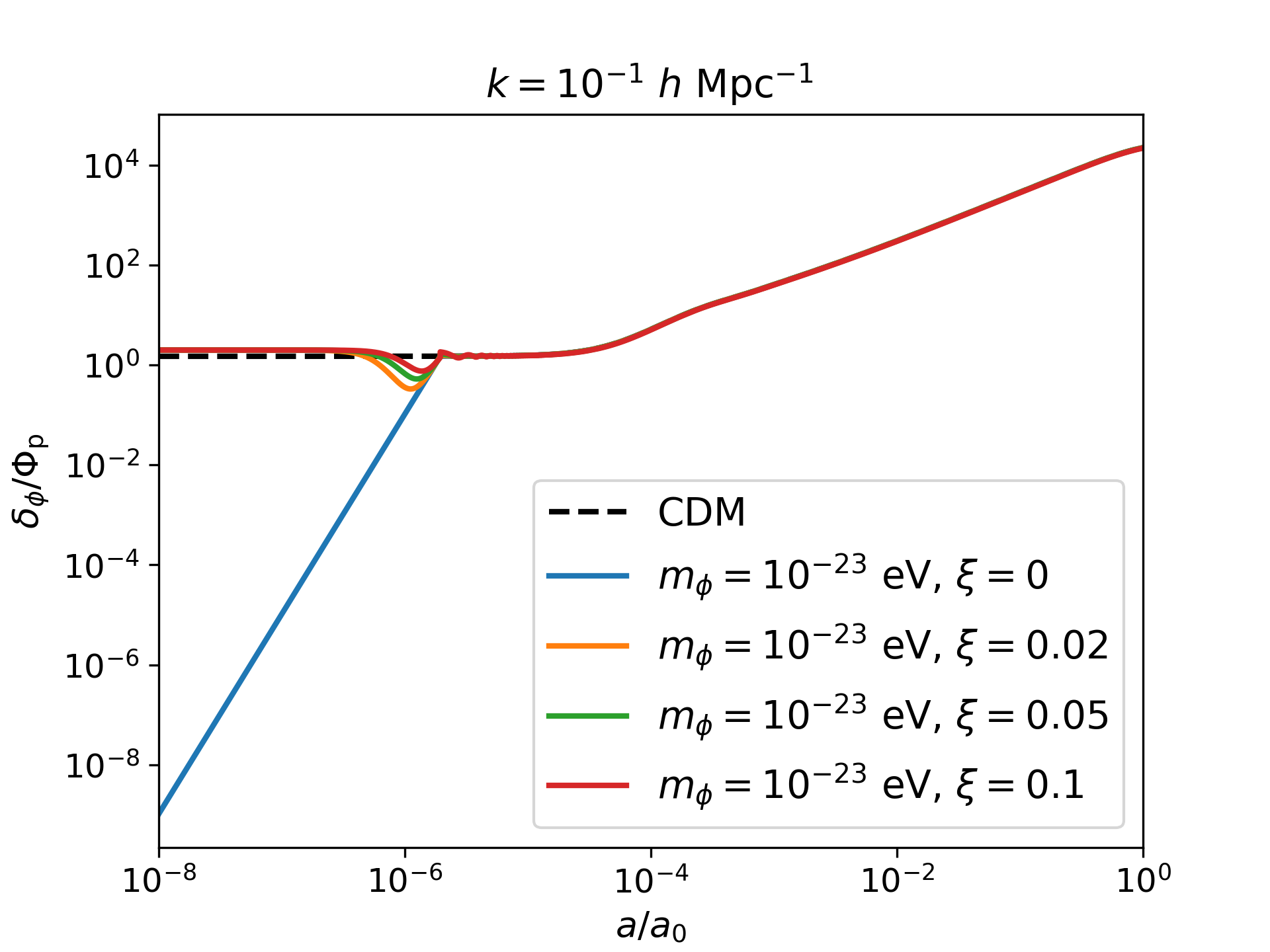}
    \end{subfigure}
    
    \caption{Evolution of ULA  density perturbation $\delta_{\phi}$ (adiabatic mode) displayed for different values of $k$. Note that     $\Phi_p$ is the primordial value of the metric perturbation $\Phi$. \label{fig:delta_phi}}
\end{figure}

To gauge the impact of nonminimal coupling, we show the evolution of the metric perturbation $\Phi$ and the scalar-field  density perturbation $\delta_{\phi}$ for the adiabatic mode for $k = 10^{-4}\ h$ Mpc$^{-1}$, $10^{-2}\ h$ Mpc$^{-1}$, and $10^{-1}\ h$ Mpc$^{-1}$ in Figs.~\ref{fig:Phi} and~\ref{fig:delta_phi}, respectively. We show four different $\xi$ values with $m_{\phi} = 10^{-23}$~eV. We also plot  the evolution of $\Phi$ and $\delta_{\rm{CDM}}$ in the  usual $\Lambda$CDM case for reference.

We first summarize the features seen in Figs.~\ref{fig:Phi} and~\ref{fig:delta_phi}: (a) The evolution of $\Phi$ for a minimally coupled scalar field  ($\xi = 0$) agrees  with the CDM case for all values of $k$. (b) After the  matter-radiation equality, the density perturbations for the minimally coupled
scalar field coincide with the CDM model. (c) For the nonminimally coupled case,  both $\Phi$ and  $\delta_{\phi}$ are suppressed after the matter-radiation equality for small $k$. The suppression is seen to scale with   $\xi$  for a given $k$.  The impact of the nonminimal coupling
decreases for smaller scales  for a fixed $\xi$. The scales that enter
the horizon before the matter-radiation equality ($k \geq 10^{-1}\ h$ Mpc$^{-1}$) are not affected appreciably by the nonminimal coupling.
(d) For a minimally coupled scalar field, the initial density perturbation $\delta_{\phi}$ is zero, whereas for the  nonminimally coupled case, it has an initial value  of  $2\Phi_{\rm p}$, the same initial condition as  for $\delta_{\gamma}$, the photon density perturbation (for further details see Appendix~\ref{sec:incon}). Thus, even at first order, a nonminimally coupled scalar field  behaves like radiation at early times.  

The main impact of nonminimal coupling is on large scales, in particular scales that enter the horizon at late times. We focus on this  issue below. 

 To understand Figs.~\ref{fig:Phi} and~\ref{fig:delta_phi}, we first consider the   evolution  of the potential,  $\Phi$,  for the $\Lambda$CDM model. In the $\Lambda$CDM model, the potential drops by a factor of 9/10  after matter-radiation equality for scales that enter the horizon deep in the matter-dominated era   ($k = 10^{-4} \ h\ \rm Mpc^{-1}$ in Fig.~\ref{fig:Phi}), while the decay is larger for smaller scales (e.g. \cite{2005pfc..book.....M,Dodelson2003}). The potential decays again when the universe becomes dominated by the cosmological constant at $z\lesssim 0.7$. 

The decay of  $\Phi$  by  a factor of 9/10 for the long wavelength modes  can be understood within the framework of  multicomponent fluids in
the $\Lambda$CDM  model.  In  the $\Lambda$CDM model, both the background and the first order
  perturbations (on superhorizon scales)  are dominated by  relativistic components,  photons and neutrinos,  in the radiation-dominated era with an effective  equation of state $w \simeq 1/3$ [Eq.~(\ref{eq:eqosta})]
  and  an adiabatic sound velocity $c_s^2 \equiv \delta p/\delta\rho \simeq  1/3$. In the
  matter-dominated era,  cold dark matter dominates the dynamics and
  both $w$ and $c_s^2$ vanish. In addition, the initial conditions
  for all components are assumed to be adiabatic, which implies that the pressure
  perturbations are entirely specified by density perturbations: $\delta p = c_s^2 \delta \rho$. Under these conditions,  it can be shown analytically
  that the initial  potential at large scales  falls by a factor of 9/10 
after the matter-radiation equality (for details, see \cite{2005pfc..book.....M}). 

In the ULA model, the background evolution of the scalar field  behaves as   a pressureless fluid, $\langle w \rangle  =0$,  for
$z \lesssim 10^5$ (Fig.~\ref{fig:bg_3}). 
However, many of the other conditions satisfied by the $\Lambda$CDM model do not hold
for this model.  In particular, the behavior of  pressure perturbations is  more complicated in this case  (Eq.~(\ref{eq:prespert})). While the pressure
perturbations become negligible for small scales for $z \lesssim 10^5$, they could be significant for large scales even after the matter-radiation equality.
In addition, for ULA models, it is not possible to express pressure perturbations
entirely in terms of density perturbations [Eqs.~(\ref{eq:denper_def})  and~(\ref{eq:prespert})]. This means the pressure perturbations in the scalar field are
not purely ``adiabatic" and the ``entropy" part of the perturbation provides an
additional source of perturbations.\footnote{Pressure perturbations of any
  component of the fluid can, in general,  be expressed in terms of two thermodynamical variables. For  isentropic initial conditions, the pressure perturbations can be written  in terms of only one variable (e.g. density perturbation),  $\delta p = c_s^2 \delta \rho$, where $c_s^2$ is the adiabatic velocity of
  sound in the medium. In principle, a whole range of other initial conditions are possible (e.g. \cite{KodamaSasaki1984} and references therein).}
An important distinction between the minimal and  nonminimal coupling is that
for the minimal coupling this split renders both  $c_s^2$ and  entropy
perturbations singular. However, for the nonminimally coupled case, the split is
well behaved (for detailed discussion, see \cite{KodamaSasaki1984}).

These additional features of the ULA model could cause the potential to
behave differently from the usual case. For a minimally coupled case, the impact of these additional features
is negligible and we notice that the behavior of $\Phi$ is similar to the CDM case. However, for the nonminimally coupled case, pressure perturbations of the scalar field  could play an important role during the transition from radiation to matter domination and  cause the potential to decay for a longer period extending into the matter domination. As the pressure perturbations are related to other perturbations through energy-momentum conservation conditions, we expect a significant impact on other components of matter perturbations also.  In particular, Fig.~\ref{fig:delta_phi} shows that the density contrast for the scalar field also decays for $k \simeq 10^{-4} h \, \rm Mpc^{-1}$. This anomalous behavior of density perturbations for large scales  results in a significant change in the observable matter power spectrum.

From  density contrasts  of different components, we can compute the matter power spectrum $P(k,t_0)$  (e.g. \cite{Dodelson2003}):
\begin{gather}
    P(k,t_0) = \frac{A}{k^3}\bigg(\frac{k}{H_0}\bigg)^{n-1}\Omega_m^2\delta_{\rm net}^2(k,t_0). \label{eq:power_spectrum}
\end{gather}
Here $\delta_{\rm net} = \sum_i f_i \delta_i$ with $f_i\equiv \Omega_i/\sum_i\Omega_i$ and $\delta_i$ being the  fraction and density contrast of different components. At the current epoch,  only the scalar field and baryons provide important contributions to the density contrast,\footnote{We assume that the only component of  dark energy is the cosmological constant for which the density contrast vanishes \cite{Ma1995}.} and we use values of $f_i$ given by the best-fit Planck parameters for baryons and CDM. Here $\Omega_m = \Omega_{\phi}(t_0) + \Omega_b(t_0)$ is the  value of the total matter density parameter at the current epoch, $A$ is a constant whose value is determined by  matching  the theoretical matter  power spectrum to cosmological observables, and $n$ is the scalar  spectral index of the initial matter power spectrum generated during the inflationary era: $P(k,t_i) \propto k^n$ with $n \simeq 1$. In this paper,  we use the best-fit value obtained by the Planck Collaboration,   $n = 0.96$ \cite{Planck2020_6}.

The matter power spectrum is probed
by the Planck  CMB data for scales in the range $k \simeq 10^{-4}\hbox{--}10^{-1} h\, \rm Mpc^{-1}$ (e.g. \cite{Planck2020_5}). Using Planck data, the matter power spectrum can be reconstructed in the aforementioned range \cite{Ade:2013zuv}.  The SDSS galaxy clustering data measure the power 
spectrum for $k \simeq 10^{-2}\hbox{--}0.4 h \, \rm Mpc^{-1}$ \cite{SDSS_1}.
We note that only the  galaxy data
for $k \lesssim 0.1 \, \rm Mpc^{-1}$ can be directly compared with the prediction of  linear perturbation theory, as linear theory fails for
smaller scales at the current epoch (e.g. \cite{Dodelson2003}). The CMB and galaxy clustering data are  compatible with  each other using the results of general relativistic perturbation theory for the usual $\Lambda$CDM model  (which includes the inflation-generated initial power-law matter power spectrum [see the  discussion following Eq.~(\ref{eq:power_spectrum}), e.g. \cite{Planck2020_5}]). One can normalize the matter power spectrum using Planck CMB temperature anisotropy and CMB lensing measurements (e.g. \cite{Ade:2013zuv}). Alternatively, it can be normalized  using the abundance of  low-redshift  massive  clusters and cosmological weak lensing data at $z \lesssim 1$ (e.g. \cite{Dodelson2003} and references therein). These data can be used to construct $\sigma_8$, the   mass dispersion at the scale $8 h^{-1} \rm Mpc$. Planck results give $\sigma_8 = 0.81$ \cite{Planck2020_5}, which is in agreement with the low redshift data.

\begin{figure}[!ht]
    \centering
    \includegraphics[width = 0.99\linewidth]{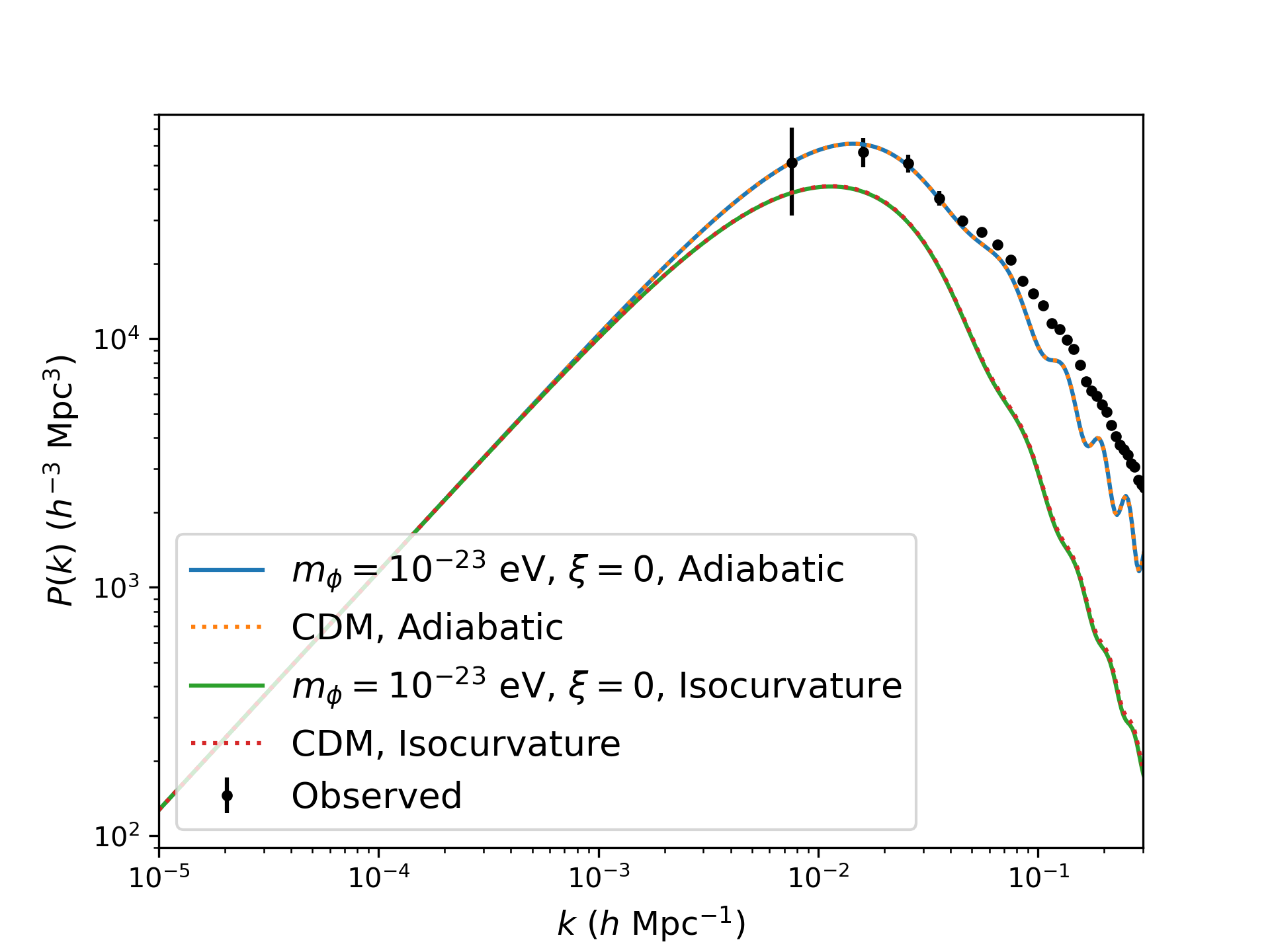}
    \caption{Adiabatic and isocurvature matter power spectra shown for minimally coupled axions for  $m_{\phi} = 10^{-23}$ eV. The $\Lambda$CDM model results are also shown for comparison. The data points corresponding to the  monopole of the measured galaxy power spectrum from SDSS are also shown (e.g. \cite{beutler2021unified}).}
    \label{fig:power_spectrum_xi=0}
\end{figure}

In Fig.~\ref{fig:power_spectrum_xi=0}, we show the matter power spectra
for $\xi =0$. Both the adiabatic and isocurvature modes\footnote{We consider only the CDM isocurvature case here, see Appendix~\ref{sec:incon} for details. This mode is shown here for the sake of  completeness as it is ruled out by the cosmological data, e.g.  \cite{Planck2020_10}} are shown for a fixed
ULA mass, $m_{\phi} = 10^{-23}$~eV.  The matter power for the two modes is matched at large scales to the $\Lambda$CDM  model. For the range of scales shown in Fig.~\ref{fig:power_spectrum_xi=0}, 
$k \lesssim 0.3 h \, \rm  Mpc^{-1}$,  the matter power spectrum for the  minimally coupled scalar field agrees  with the  power spectrum for the usual $\Lambda$CDM case. This inference is compatible  with existing results in the literature (e.g. \cite{UrenaLopez2016}).\footnote{One can  compute the minimally-coupled ULA matter power spectra using the code \textit{AxionCAMB} which
 is a modified version of the
publicly-available code CAMB; it is available at https://github.com/dgrin1/axionCAMB} We note that, for the case shown in Fig.~\ref{fig:power_spectrum_xi=0}, the ULA matter power spectrum has  a  power deficit  as compared to
the $\Lambda$CDM model at  smaller scales,  $k \gtrsim 1$ Mpc$^{-1}$ (these scales are not shown in the figure),  and multiple studies have considered the implications of this  small-scale matter  power suppression\footnote{For smaller $m_\phi$, the matter power spectrum deviates from the $\Lambda$CDM model for
  larger scales; for instance, for $m_\phi = 10^{-24} \, \rm eV$, the suppression occurs for $k \gtrsim 0.3 \, \rm Mpc^{-1}$. An approximate relation between this scale and
  the scalar-field mass can be obtained by showing that the scalar field
  behaves as a medium with a time- and scale-dependent effective sound speed   at subhorizon scales (e.g. see \cite{Marsh2016}  for details). In this paper, we only present results for $m_\phi = 10^{-23} \, \rm eV$, but we verified  the expected  small-scale behavior for ULA models for smaller  masses.} \cite{frieman1995cosmology,coble1997dynamical,hu2000fuzzy,marsh2010ultralight,park2012axion,kobayashi2017lyman, Hui:2016ltb,2016JCAP...04..012S,2017JCAP...07..012S,Hlozek2015,rogers2020strong}.  However,  our focus in this paper is on  large scales.
 
\begin{figure}[!ht]
    \centering
    \includegraphics[width = 0.99\linewidth]{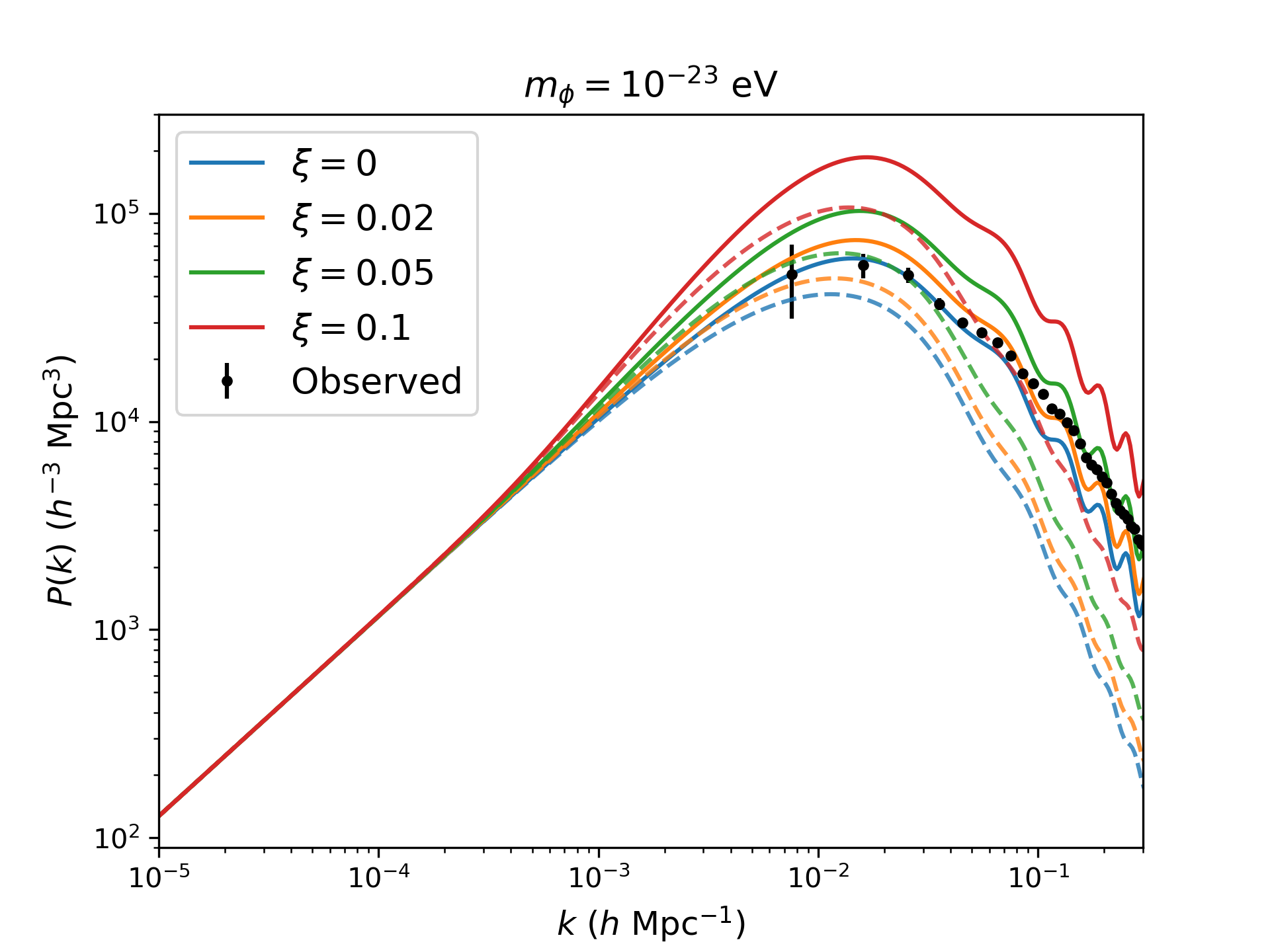}
    \caption{Adiabatic and isocurvature matter power spectra for the nonminimally coupled ULAs shown  for different values of $\xi$. The solid lines denote  the adiabatic power spectra and the dashed lines show the isocurvature power spectra.}
    \label{fig:power_spectrum_nmc}
\end{figure}

In Fig.~\ref{fig:power_spectrum_nmc}, we display the matter power spectrum for the
nonminimally coupled case for $m_{\phi} = 10^{-23}$~eV.
As discussed above, in this case, the density perturbations in the scalar field   deviate  from the CDM density perturbations  at large scales but reproduce the results of the CDM  model at small scales. This leaves observable traces on the observed matter power spectrum. We match  the  matter power spectrum to the $\Lambda$CDM model   at large scales. 
Given the decay of density perturbations  on large scales for the nonminimally coupled scalar field   (Fig.~\ref{fig:delta_phi}), the matter power is expected to  differ  between the nonminimal coupling case and the  $\Lambda$CDM (or minimal coupling case) at small scales.  As a result, the matter power is found to be
significantly larger at small scales as compared to  the CDM power spectrum. Alternatively, if we had chosen to normalize the power 
spectrum at small scales (e.g. by choosing the measured value of $\sigma_8$ by cluster abundance data), we would find a deficit of 
matter power at large scales. 
The power excess in Fig.~\ref{fig:power_spectrum_nmc} is proportional to the value of $\xi$. To obtain agreement with both the large- and 
small-scale data, we obtain the following 
constraint  on the nonminimal coupling for  the  adiabatic mode: $\xi \lesssim 0.01$.

Figure~\ref{fig:power_spectrum_nmc} also shows that, for nonminimal coupling, the matter power spectrum for  the isocurvature mode   is in much better agreement with   the
usual adiabatic CDM mode. For instance, the isocurvature mode for $\xi = 0.05$
agrees well  with the $\xi = 0$ adiabatic mode at large scales. We do not carry out
a more detailed comparison here but our results  suggest  that a mix
of adiabatic and isocurvature initial conditions  for the nonminimally coupled ULAs might
 behave similarly to  the  CDM  adiabatic mode. 

In Figs.~\ref{fig:Phi},~\ref{fig:delta_phi}, and~\ref{fig:power_spectrum_nmc}, we show results for a fixed  ULA mass $m_\phi = 10^{-23} \, \rm eV$  as a function of the nonminimal coupling. For our work we consider
  a mass range $10^{-25} \, {\rm eV} < m_\phi < 10^{-19} \, \rm eV$. We note that
  the large-scale behavior  shown in  the figures is obtained for this entire range of masses. The best reported upper limit on the mass $m_\phi \gtrsim 10^{-20} \, \rm eV$ arises from the small-scale behavior of the power spectrum using  the Lyman-$\alpha$ data \cite{rogers2020strong}. Our main results are compatible with this upper limit.

\section{Conclusion \label{sec:concl}}

In this paper, we study a nonminimally coupled scalar-field
as a potential cold dark matter candidate. The minimal scalar field
models have been extensively  studied as  ultralight axions in the literature
and are known to alleviate well-known small-scale issues with  the usual
$\Lambda$CDM model. Our study is a natural extension of such models.

The dynamics of both  the background and the perturbed  components of the scalar field change substantially   for the nonminimally coupled case.  Initially, the background scalar field  behaves as radiation,
unlike the usual case in which the scalar-field energy starts as a cosmological
constant. For a scale factor in the range $a \simeq 10^{-5}\hbox{--}10^{-7}$, the nonminimally coupled scalar
field makes a transition to the cosmological-constant domination phase (Fig.~\ref{fig:background}). The altered initial
evolution yields a new constraint on such models from primordial nucleosynthesis. During primordial nucleosynthesis, the final
abundance of helium-4 and deuterium
is a sensitive function of the total radiation content of
the universe at $a \simeq 10^{-9}$. A comparison of current data with
the theory of formation of light elements suggest that the amount of
``dark radiation" could not exceed  10\% of the radiation energy density
given by photons and the standard model neutrinos during that era (e.g. see  \cite{2012arXiv1208.0032S} and references therein). This constrains the 
strength of the gravity-scalar-field coupling $\xi \lesssim 10$ for the range of scalar-field masses we consider here.

The first order perturbation theory of a nonminimally coupled  scalar field
adds multiple new complexities as compared to the usual case. One of the
new features is the presence of anisotropic stress, which mainly impacts 
small scales [Eq.~(\ref{eq:axion_anis_stress})]. Our most important finding in this paper   relates to large scales. For nonzero $\xi$, the perturbations on  scales
that enter the horizon after the matter-radiation equality could have  radically different 
behavior (Figs.~\ref{fig:Phi} and~\ref{fig:delta_phi}).  This causes the
matter power spectrum to deviate significantly from the minimal coupling case
(Fig.~\ref{fig:power_spectrum_nmc}). For adiabatic initial conditions, a comparison of the computed matter power spectrum
with galaxy clustering and CMB anisotropy data puts strong constraints on the nonminimal
coupling: $\xi \lesssim 0.01$. 

We also consider isocurvature  initial conditions. More specifically,
we consider the scalar-field isocurvature mode in which only the
scalar-field density contrast is nonzero initially. 
Figure~\ref{fig:power_spectrum_nmc} shows that this isocurvature mode
for nonzero $\xi$ agrees with the adiabatic CDM (or minimally coupled) mode
at large scales. This means that a mix of isocurvature and adiabatic
initial conditions  might explain the observed matter power spectrum
for a nonminimally coupled ULA.

In the current work, we focus on computing the matter power spectrum for nonminimally coupled ULAs and compare the prediction of the model with the galaxy power spectrum and CMB results.
This yields an approximate bound on the strength of nonminimal coupling. A more detailed multiparameter analysis based on either the Fisher matrix or MCMC methods will give more precise constraints.   This
work also needs to be extended to CMB temperature and polarization anisotropies, as a   direct comparison with the CMB anisotropy data
would help quantify our results further. In particular, the  new physics our model introduces   might result in  novel outcomes such as the following:  (a) the integrated Sachs-Wolfe effect would be altered owing to  the time dependence of the potential at large scales, (b)  the sound horizon close to the era of recombination might change such that it would have a bearing on the issue of Hubble tension between the CMB and low-redshift data (e.g. \cite{2021CQGra..38o3001D} and references therein).  We hope to
return to these issues with the theoretical computation of  the CMB temperature and polarization anisotropies and a more detailed statistical comparison with cosmological data in a future work.

\section{Acknowledgement}

We thank Dr. Florian Beutler for providing us with the data points for the observed galaxy  power spectrum.

\appendix

\appendix

\section{EINSTEIN AND FLUID EQUATIONS} \label{sec:eineq}

In this section, we give the necessary cosmological equations, including  Einstein equations and the relevant  equations for photons, baryons, neutrinos, and cosmological constant at zeroth and first order (for details, see e.g. \cite{Bardeen1980, KodamaSasaki1984, Ma1995, Dodelson2003, UrenaLopez2016}).

\subsection{Background equations \label{sec:Appendix_Background_Equations}}

 For the sake of completeness and
consistency of notations, we list the relevant equations for the background evolution of the universe.

We have already defined the scalar field, photon, and neutrino density parameters in Sec. \ref{sec:Background_Equations}.
We can similarly define the density parameter for any component of the universe by
\begin{equation}
    \Omega_{i} = \frac{8\pi G\rho_i}{3H^2}. \label{eq:density_parameter_defn}
\end{equation}
We denote the baryon density parameter by $\Omega_b$ and the cosmological-constant density parameter by $\Omega_{\Lambda}$.

In general, the density parameter for the $i$th component evolves as
\begin{equation}
    \Omega_i' = 3(w - w_i)\Omega_i, \label{eq:dens_parameter_evolution}
\end{equation}
where $w_i$ is the equation of state of the $i$th component. Thus, the density parameters evolve according to the following equations:
\begin{gather}
    \Omega_{\gamma}' = 3\bigg(w - \frac{1}{3}\bigg)\Omega_{\gamma}, \label{eq:Omega_gamma_prime} \\
    \Omega_{\nu}' = 3\bigg(w - \frac{1}{3}\bigg)\Omega_{\nu}, \label{eq:Omega_nu_prime} \\
    \Omega_b' = 3w\Omega_{b}, \label{eq:Omega_b_prime} \\
    \Omega_{\Lambda}' = 3(w + 1)\Omega_{\Lambda}.\label{eq:Omega_Lambda_prime}
\end{gather}
Using Eq.~(\ref{eq:eqosta}), these equations specify the evolution of all 
the density parameters. 
In addition to these fluid equations, we have two Einstein equations given by
\begin{gather}
    H^2 = \frac{\Dot{a}^2}{a^4} = \frac{8\pi G}{3}\sum_i \rho_i, \label{eq:Einstein_1} \\
    \Dot{H} = -4\pi Ga \sum_i (\rho_i + p_i). \label{eq:Einstein_2}
\end{gather}
Equation~(\ref{eq:Einstein_1}) can be written in terms of the density parameters as
\begin{gather}
    \Omega_{\gamma} + \Omega_{\nu} + \Omega_{\phi} + \Omega_b + \Omega_{\Lambda} = 1. \label{eq:Algebraic_Einstein_eqn}
\end{gather}
On the other hand, Eq.~(\ref{eq:Einstein_2}) converts to an equation for $y$, defined in Sec. \ref{sec:Background_Equations} as $y = 2m_{\phi}/H$. We obtain
\begin{gather}
    y' = \frac{3}{2}(1 + w)y. \label{eq:y_prime}
\end{gather}

Equations~(\ref{eq:Omega_phi_prime})--(\ref{eq:w_phi_defn}), along with Eqs.~(\ref{eq:Omega_gamma_prime})--(\ref{eq:Omega_Lambda_prime}) and~(\ref{eq:Algebraic_Einstein_eqn})--(\ref{eq:y_prime}), constitute the complete set of background equations. 

\subsection{First order equations \label{sec:Appendix_First_Order_Equations}}

In Fourier space, the first order equations for photons, neutrinos, and baryons are (for details, see e.g. \cite{Bardeen1980, KodamaSasaki1984, Ma1995, Dodelson2003})
\begin{gather}
    \Theta_0' + \frac{k}{aH}\Theta_1 = -\Phi', \label{eq:Theta_0_prime}\\
    \Theta_1' - \frac{k}{3aH}\Theta_0 = \frac{k}{3aH}\Psi - \frac{n_e\sigma_T}{H}\bigg(\Theta_1 - \frac{iv_b}{3}\bigg),\\
    \mathcal{N}_0' + \frac{k}{aH}\mathcal{N}_1 = -\Phi',\\
    \mathcal{N}_1' - \frac{k}{3aH}\left (\mathcal{N}_0 -\mathcal{N}_2\right) = \frac{k}{3aH}\Psi,\\
    \delta_b' + \frac{ik}{aH}v_b = -3\Phi',\\
    v_b' + v_b = -\frac{ik}{aH}\Psi - \frac{n_e\sigma_T}{HR}(v_b + 3i\Theta_1).
\end{gather}
Our notation is consistent with that of Dodelson \cite{Dodelson2003}. Here $\Theta_0$ and $\Theta_1$ are the monopole and the dipole components of the photon temperature perturbation, respectively; $\mathcal{N}_0$ and $\mathcal{N}_1$ are the monopole and dipole components of the neutrino temperature perturbation, and $\mathcal{N}_2$ is the quadrupole moment. The quadrupole moment for photons
is omitted as it is negligible in the tight-coupling approximation \cite{Dodelson2003}. Note that $\delta_b = \delta\rho_b/\rho_b$ is the baryon overdensity, $v_b$ is the bulk  velocity of baryons, $n_e$ is the electron number density, $\sigma_T$ is the Thomson scattering cross section, and $R = 3\Omega_b/4\Omega_{\gamma}$ is the baryon-energy-to-photon-energy ratio.

We simplify photon  equations
further in the tight-coupling approximation and retain terms up to the first order in the scattering
timescale, $1/(n_e\sigma_T c)$. This procedure neglects Silk damping, which
is needed for  accurate treatment of CMB anisotropies and the matter power spectrum at small scales. However, it
makes a negligible impact on our treatment  as the main impact of our results is at large scales. For massless neutrinos, we adopt $\mathcal{N}_2 = 0$ and  neglect  higher multipoles. These  modes play an
important role for  subhorizon modes,  $k\eta > 1$. For $\ell \ge 2$, these
modes decay as $j_\ell(k\eta)$ after the horizon entry (e.g. Ma and Bertschinger \cite{Ma1995}). We check the efficacy
of our procedure by putting  neutrino perturbation to zero after the horizon
entry, and we find that a more precise treatment   of higher neutrino multipoles  does not  significantly affect the predicted matter power spectrum on the scales of interest to us.

The perturbed metric components, $\Phi$ and $\Psi$, obey the following first order Einstein equations.
\begin{multline}
  \Phi' - \Psi + \frac{k^2}{3a^2H^2}\Phi = \frac{1}{2}(4\Omega_{\gamma}\Theta_0 + 4\Omega_{\nu}\mathcal{N}_0 \\ + \Omega_{\phi}\delta_{\phi} + \Omega_b\delta_b),
  \label{eq:einfir}
\end{multline}
\begin{multline}
    \Phi' - \Psi = -\frac{3}{2}\frac{aH}{k}[4\Omega_{\gamma}\Theta_1 + 4\Omega_{\nu}\mathcal{N}_1 \\ +  \Omega_{\phi}(1+w_{\phi})v_{\phi} + \Omega_bv_b],
\end{multline}
\begin{multline}
    \Phi'' + 3\Phi' - \Psi' - 3\Psi - \frac{3}{2}(1 + w)(\Phi' - 2\Psi) + \\ \frac{k^2}{3a^2H^2}(\Phi + \Psi) = -\frac{3}{2}\bigg(\frac{4}{3}\Omega_{\gamma}\Theta_0 + \frac{4}{3}\Omega_{\nu}\mathcal{N}_0 + \Omega_{\phi}w_{\phi}\pi_{\phi}\bigg),
\end{multline}
\begin{gather}
    -\frac{k^2}{3a^2H^2}(\Phi + \Psi) = \Omega_{\phi}w_{\phi}\Pi_{\phi}. \label{eq:Anisotropic_stress_equation}
\end{gather}
Equations~(\ref{eq:alpha_eqn})--(\ref{eq:axion_anis_stress}), along with Eqs.~(\ref{eq:Theta_0_prime})--(\ref{eq:Anisotropic_stress_equation}), constitute the complete set of first order equations. Not all of these first order equations are independent of each other. In particular, we solve two Einstein's equations [Eqs.~(\ref{eq:einfir}) and~(\ref{eq:Anisotropic_stress_equation})] along with the fluid
equations. The other two Einstein's equations can be  derived from the
equations we use (for details, see e.g. Kodama and Sasaki \cite{KodamaSasaki1984}).

\section{INITIAL CONDITIONS} \label{sec:incon}

To obtain the initial conditions, we analytically solve the set of background and first order equations  in the radiation-dominated era such that  $k\eta \ll 1$ for all scales of interest to us. However, for nonminimally coupled scalar
fields, the complex nature of equations does not always permit an analytic solution
even at early times. In such cases, we choose the initial condition for   $\xi = 0$ and numerically search for the suitable  initial conditions for the nonminimally coupled case. We assume that the correct solution  is an attractor solution and the approximate initial conditions allow us to reach the relevant solution 
once the nonminimally coupled field enters the cosmological-constant-dominated phase. 

We follow the work of Ure{\~n}a-L{\'o}pez and Gonzalez-Morales \cite{UrenaLopez2016} to obtain the initial conditions for variables related to ULAs. For initial conditions for other variables, we follow the work of Dodelson \cite{Dodelson2003}.

\subsection{Background initial conditions \label{sec:Background_Initial_Conditions}}

We find the initial conditions at the scale factor $a = a_i$ such that $m_{\phi}/H_i \ll 1$, that is, $y_i \ll 1$. This is expected deep in  the radiation-dominated epoch. Thus, we expect $\Omega_{\gamma i} + \Omega_{\nu i} \simeq 1$, and $\Omega_{ai} \ll 1$ for $a \neq \gamma,\ \nu$. The initial condition for the scalar field is chosen such that $\Dot{\phi}_i = 0$ and $\phi_i < 0$. Therefore,  the initial  value of $\theta$  satisfies the condition: $\theta_i \ll 1$ [Eqs.~(\ref{eq:transformation1}) and~(\ref{eq:transformation2})].

Using these approximations, we have, at early times, $H = H_0\sqrt{\Omega_{\gamma0} + \Omega_{\nu0}}/a^2$. Thus, we have the initial condition for $y$,
\begin{gather}
    y_i = \frac{2m_{\phi}}{H_i} = \frac{2m_{\phi}a_i^2}{H_0\sqrt{\Omega_{\gamma0} + \Omega_{\nu0}}}. \label{eq:y_init_cond}
\end{gather}

Both Eqs.~(\ref{eq:Omega_phi_prime}) and~(\ref{eq:theta_prime}) contain a term $(1 - 3w)/y$. This term can be approximated at early times as follows.
\begin{align}
    1 - 3w &= 1 - (\Omega_{\gamma} + \Omega_{\nu} + \Omega_{\phi}) + 3\Omega_{\Lambda} = \Omega_b + 4\Omega_{\Lambda}.
\end{align}
Here, we have used Eq.~(\ref{eq:Algebraic_Einstein_eqn}) and the fact that, at early times, $w_{\phi} = 1/3$ [see Eq.~(\ref{eq:w_phi_approximation})]. From Eq.~(\ref{eq:density_parameter_defn}), we have $\Omega_{b}/(\Omega_{\gamma} + \Omega_{\nu}) = \rho_b/(\rho_{\gamma} + \rho_{\nu}) = a\rho_{b0}/(\rho_{\gamma0} + \rho_{\nu0})$. Thus, the initial condition for $\Omega_b$ is
\begin{gather}
    \Omega_{bi} = \frac{\Omega_{b0}a_i}{\Omega_{\gamma0} + \Omega_{\nu0}}.
\end{gather}
Similarly,
\begin{gather}
    \Omega_{\Lambda i} = \frac{\Omega_{\Lambda 0}a_i^4}{\Omega_{\gamma0} + \Omega_{\nu0}}.
\end{gather}
So, we obtain
\begin{gather}
    \frac{1 - 3w}{y} = \frac{H_0}{2m_{\phi}\sqrt{\Omega_{\gamma0} + \Omega_{\nu0}}}\bigg(\frac{\Omega_{b0}}{a} + 4\Omega_{\Lambda 0}a^2\bigg)
\end{gather}
deep in the radiation-dominated epoch.

Finally, using the approximation $\theta_i \ll 1$, Eq.~(\ref{eq:theta_prime}) gives
\begin{multline}
    \theta' = -3\theta + \frac{2m_{\phi}a^2}{H_0\sqrt{\Omega_{\gamma0} + \Omega_{\nu0}}} \\ + \frac{6\xi H_0}{m_{\phi}\sqrt{\Omega_{\gamma0} + \Omega_{\nu0}}}\bigg(\frac{\Omega_{b0}}{a} + 4\Omega_{\Lambda 0}a^2\bigg).
\end{multline}
The solution to this ODE is
\begin{multline}
    \theta = \frac{2}{5\sqrt{\Omega_{\gamma0} + \Omega_{\nu0}}}\bigg(\frac{m_{\phi}}{H_0} + \frac{12\xi H_0\Omega_{\Lambda0}}{m_{\phi}}\bigg)a^2 \\ + \frac{3\xi H_0\Omega_{b0}}{m_{\phi}\sqrt{\Omega_{\gamma0} + \Omega_{\nu0}}}\frac{1}{a} + \frac{C}{a^3},
\end{multline}
where $C$ is the constant of integration. We neglect the fastest decaying mode, which is  the last term in the  solution. This gives the initial condition for $\theta$,
\begin{multline}
    \theta_i = \frac{2}{5\sqrt{\Omega_{\gamma0} + \Omega_{\nu0}}}\bigg(\frac{m_{\phi}}{H_0} + \frac{12\xi H_0\Omega_{\Lambda0}}{m_{\phi}}\bigg)a_i^2 \\ + \frac{3\xi H_0\Omega_{b0}}{m_{\phi}\sqrt{\Omega_{\gamma0} + \Omega_{\nu0}}}\frac{1}{a_i}. \label{eq:theta_init_cond}
\end{multline}

Obtaining the initial condition for $\widetilde{\Omega}_{\phi}$ in this manner is difficult. Thus, we put $\xi = 0$ and follow the method used in the work of Ure{\~n}a-L{\'o}pez and Gonzalez-Morales \cite{UrenaLopez2016}. This gives the following initial condition for $\widetilde{\Omega}_{\phi}$:
\begin{gather}
    \widetilde{\Omega}_{\phi i} = K\frac{\Omega_{\phi 0}}{\Omega_{\gamma0} + \Omega_{\nu0}}a_i\bigg[\frac{4\theta_i^2}{\pi^2}\bigg(\frac{9 + \pi^2/4}{9 + \theta_i^2}\bigg)\bigg]^{3/4},
\end{gather}
where $K$ is a constant that needs to be found numerically. We can use a binary search in order to find the value of $K$ which gives the correct value of $\Omega_{\phi0}$, the present value of scalar-field density parameter. For the range of cases we consider for the  nonminimally coupled ULAs, the value of $K$ can vary by many orders of magnitude.

The initial condition for $\Omega_{\nu}$ is  given by
\begin{gather}
    \Omega_{\nu i} = \frac{\Omega_{\nu0}}{\Omega_{\gamma0} + \Omega_{\nu0}}.
\end{gather}

\subsection{First order initial conditions \label{sec:Firstorder_Initial_Conditions}}

For the first order equations, we consider two  initial conditions: adiabatic and isocurvature. In a multicomponent fluid, these initial conditions can
refer to different components. We adopt the  usual  adiabatic initial conditions in which  the ratio of the number densities of different components is chosen to be the  same  (for a detailed discussion,
see e.g. \cite{Ma1995} or \cite{2005pfc..book.....M}). On the other hand, for isocurvature initial conditions, only the density perturbations associated with either the baryons or ULAs are  nonzero initially. For instance, the ULA isocurvature mode has  only $\delta_{\phi} \neq 0$ during very early times, while all other perturbation variables are set to zero. We consider only this isocurvature
mode in our analysis. 

Obtaining initial conditions at first order using analytic solutions for nonminimally coupled scalar fields is difficult. Therefore, as in the background case,  we start our search for  the suitable  initial conditions with  the $\xi = 0$ case for both modes. 

\subsubsection{Adiabatic mode}

Using Eqs.~(\ref{eq:axion_anis_stress}) and~(\ref{eq:Anisotropic_stress_equation}), we obtain $\Psi = -\Phi$ for $\xi = 0$. For the  adiabatic mode, the metric perturbation is constant deep in  the radiation-dominated era. Therefore, $\Phi_i' = 0$. Further, we know from the previous section that as $a\to0$, $y\to0$ and $\theta\to0$. Thus, at early times, from Eq. (\ref{eq:beta_eqn}), we obtain
\begin{gather}
    \lim_{a\to0}\beta' = -3\sin(\theta-\beta) - \frac{k^2}{k_J^2}[1-\cos(\theta-\beta)].
\end{gather}
Note that $\theta = \beta$ is a critical point of the system. Substituting this value in Eq.~(\ref{eq:alpha_eqn}), we obtain:
\begin{gather}
    \alpha' = -\frac{3}{2}\alpha(1 + \cos\theta) + 2y\cos\bigg(\frac{\theta}{2}\bigg)\Phi.
\end{gather}
Considering $\theta$ to be negligible, we have
\begin{gather}
    \alpha' = -3\alpha + 2y\Phi. \label{eq:alpha_init_cond_intermediate}
\end{gather}
From Eqs.~(\ref{eq:y_init_cond}) and~(\ref{eq:theta_init_cond}), we see that, at early times, $y = 5\theta = 5\theta_i(a/a_i)^2$ for $\xi = 0$. Substituting this value of $y$ in Eq.~(\ref{eq:alpha_init_cond_intermediate}) and solving the resultant ODE, we obtain the following initial condition for $\alpha$:
\begin{gather}
    \alpha_i = 2\theta_i\Phi_i.
\end{gather}
Here we have neglected a decaying mode.

Substituting the initial condition for $\alpha$ in Eq.~(\ref{eq:beta_eqn}), we get
\begin{gather}
    \beta' = -6\theta + 3\beta + \frac{y}{\theta}(\theta - \beta) = -2\beta - \theta
\end{gather}
for the $\xi = 0$ case. Thus,
\begin{gather}
    \beta' = -2\beta - \theta_i\bigg(\frac{a}{a_i}\bigg)^2.
\end{gather}
Solving this equation, and neglecting the decaying mode, we have
\begin{gather}
    \beta_i = -\frac{\theta_i}{4}.
\end{gather}

The initial conditions for other variables are given in the work of Dodelson \cite{Dodelson2003}:
\begin{gather}
    3\Theta_{0i} = 3\mathcal{N}_{0i} = \delta_{bi} = \frac{3}{2}\Phi_i, \\
    3\Theta_{1i} = 3\mathcal{N}_{1i} = iv_{bi} = -\frac{k}{2aH}\Phi_i.
\end{gather}
Since $k/aH$ is negligible at very early times, $\Theta_{1i} = \mathcal{N}_{1i} = v_{bi} \approx 0$.

\subsubsection{ULA isocurvature mode}

For the ULA isocurvature case, we choose $\alpha_i = 1$  and $\beta_i = -\pi$.
This gives  us $\delta_{\phi i} = 1$ [see Eq.~(\ref{eq:denper_def})]. 
All the  other perturbation variables  are  set to zero initially, including the metric perturbations, $\Phi$ and $\Psi$.

\section{NUMERICAL IMPLEMENTATION} \label{sec:numimpl}
In this section, we briefly summarize the salient aspects of  numerical
implementation of the coupled Einstein-Boltzmann equations along with the scalar-field dynamics. We perform numerical integration of the relevant equations using PYTHON codes.

The background  and first order initial conditions are set using the procedures
described in Appendixes \ref{sec:Background_Initial_Conditions}  and \ref{sec:Firstorder_Initial_Conditions}. We implement the initial conditions to the lowest order
in $k\eta$ for the relevant variables, which means zeroth
order for potentials and the density field, first order for bulk velocities, and
second order for the  anisotropic stress (for details, see e.g. \cite{Ma1995}).
Using matrix methods, one can systematically develop initial conditions to
higher orders in $k\eta$ \cite{2013MNRAS.434.1619C,2018PhRvD..97f3511M,2003PhRvD..68f3505D,2018MNRAS.476.3063H,Hlozek2015}, which is harder for us to implement as the initial conditions for the ULA  have to be determined numerically in our case.

We verify the robustness of our
initial conditions by checking that   the final results do not depend on 
the choice of starting time. One novel initial condition in our case corresponds to the early-time radiationlike  behavior of the nonminimally coupled scalar field.  In all the cases we studied, the field makes a transition to the cosmological-constant phase before the onset of the oscillatory phase.  To check the  numerical stability of the initial conditions, we  slowly switch  off the
nonminimal coupling and verify that we obtain the relevant results for the minimal coupling case. An additional numerical issue for the ULA models is
to establish the smooth transition from the oscillatory phase to the time-averaged phase. As noted above, we follow the procedure given by the numerical implementation
of \cite{UrenaLopez2016}  and find a satisfactory outcome.

\bibliography{apssamp}

\end{document}